\documentclass[journal]{IEEEtran}

\usepackage{amsmath,graphicx}

\usepackage{cite}

\usepackage{amssymb}

\usepackage{array}
\usepackage{url}

\def\kron{\otimes}
\def\tr{\mathrm{tr}}

\def\diag{\mathrm{diag}}

\DeclareMathOperator*{\Htran}{\scriptscriptstyle{H}}
\DeclareMathOperator*{\Ttran}{\scriptscriptstyle{T}}

\newcommand{\condSum}[3]{\overset{#3}{\underset{\underset{#2}{#1}}{\sum}}}

\newcommand{\fracSumtwo}[2]{\overset{#2}{\underset{#1}{\sum}}}
\newcommand{\vect}[1]{\mathbf{#1}}

\newtheorem{theorem}{Theorem}
\newtheorem{corollary}{Corollary}
\newtheorem{lemma}{Lemma}

\newtheorem{example}{Example}

\begin{document}

\title{Massive MIMO with Non-Ideal Arbitrary Arrays: Hardware Scaling Laws and Circuit-Aware Design}

\author{Emil~Bj\"ornson,~\IEEEmembership{Member,~IEEE,}
        Michail Matthaiou,~\IEEEmembership{Senior~Member,~IEEE,}
        and~M\'erouane~Debbah,~\IEEEmembership{Fellow,~IEEE}%
\thanks{\copyright 2015 IEEE. Personal use of this material is permitted. Permission from IEEE must be obtained for all other uses, in any current or future media, including reprinting/republishing this material for advertising or promotional purposes, creating new collective works, for resale or redistribution to servers or lists, or reuse of any copyrighted component of this work in other works.}%
\thanks{Manuscript received July 4, 2014; revised November 3, 2014 and
February 16, 2015; accepted March 21, 2015. This research has received
funding from the EU 7th Framework Programme under GA no ICT-619086
(MAMMOET). This research has been supported by ELLIIT, the International
Postdoc Grant 2012-228 from the Swedish Research Council and the ERC
Starting Grant 305123 MORE (Advanced Mathematical Tools for Complex
Network Engineering). The associate editor coordinating the review of this
paper and approving it for publication was G.Yue.}%
\thanks{E.~Bj\"ornson  was with the KTH Royal Institute of Technology, Stockholm, SE 100 44, Sweden, and with Sup\'elec, Gif-sur-Yvette 91191, France. He is now with the Department of Electrical Engineering (ISY), Link\"{o}ping University, Link\"{o}ping, SE 581 83, Sweden (e-mail: emil.bjornson@liu.se).}%
\thanks{M.~Matthaiou is with the School of Electronics, Electrical Engineering and Computer Science, Queen's University Belfast, Belfast BT7 1NN, U.K., and also with the Department of Signals and Systems, Chalmers University of Technology, Gothenburg, SE 412 96, Sweden (e-mail: m.matthaiou@qub.ac.uk).}%
\thanks{M.~Debbah is CentraleSupelec, Gif-sur-Yvette 91191, France (email: merouane.debbah@centralesupelec.fr).}%
\thanks{Digital Object Identifier 10.1109/TWC.2015.2420095}}

\markboth{IEEE Transactions on Wireless Communications}%
{IEEE Transactions on Wireless Communications}

\maketitle

\begin{abstract}
Massive multiple-input multiple-output (MIMO) systems are cellular networks where the base stations (BSs) are equipped with unconventionally many antennas, deployed on co-located or distributed arrays. Huge spatial degrees-of-freedom are achieved by coherent processing over these massive arrays, which provide strong signal gains, resilience to imperfect channel knowledge, and low interference. This comes at the price of more infrastructure; the hardware cost and circuit power consumption scale linearly/affinely with the number of BS antennas $N$. Hence, the key to cost-efficient deployment of large arrays is low-cost antenna branches with low circuit power, in contrast to today's conventional expensive and power-hungry BS antenna branches. Such low-cost transceivers are prone to hardware imperfections, but it has been conjectured that the huge degrees-of-freedom would bring robustness to such imperfections. We prove this claim for a generalized uplink system with multiplicative phase-drifts, additive distortion noise, and noise amplification. Specifically, we derive closed-form expressions for the user rates and a scaling law that shows how fast the hardware imperfections can increase with $N$ while maintaining high rates. The connection between this scaling law and the power consumption of different transceiver circuits is rigorously exemplified. This reveals that one can make the circuit power increase as $\sqrt{N}$, instead of linearly, by careful circuit-aware system design.
\end{abstract}


\section{Introduction}

Interference coordination is the major limiting factor in cellular networks, but modern multi-antenna base stations (BSs) can control the interference in the spatial domain by coordinated multipoint (CoMP) techniques \cite{Karakayali2006a,Gesbert2010a,Bjornson2013d}.  The cellular networks are continuously evolving to keep up with the rapidly increasing demand for wireless connectivity \cite{Holma2011a}. Massive densification, in terms of more service antennas per unit area, has been identified as a key to higher area throughput in future wireless networks \cite{Hoydis2013c,Baldemair2013a,Larsson2014a}. The downside of densification is that even stricter requirements on the interference coordination need to be imposed. Densification can be achieved by adding more antennas to the macro BSs and/or distributing the antennas by ultra-dense operator-deployment of small BSs. These two approaches are non-conflicting and represent the two extremes of the \emph{massive MIMO} paradigm \cite{Larsson2014a}: a large co-located antenna array or a geographically distributed array (e.g., using a cloud RAN approach \cite{CRAN2011}). The massive MIMO topology originates from \cite{Marzetta2010a} and has been given many alternative names; for example, large-scale antenna systems (LSAS), very large MIMO, and large-scale multi-user MIMO. The main characteristics of massive MIMO are that each cell performs coherent processing on an array of hundreds (or even thousands) of active antennas, while simultaneously serving tens (or even hundreds) of users  in the uplink and downlink. In other words, the number of antennas, $N$, and number of users per BS, $K$, are unconventionally large, but differ by a factor two, four, or even an order of magnitude. For this reason, massive MIMO brings unprecedented spatial degrees-of-freedom, which enable strong signal gains from coherent reception/transmit beamforming, give nearly orthogonal user channels, and resilience to imperfect channel knowledge \cite{Rusek2013a}.

Apart from achieving high area throughput, recent works have investigated additional ways to capitalize on the huge degrees-of-freedom offered by massive MIMO. Towards this end, \cite{Hoydis2013c} showed that massive MIMO enables fully distributed coordination between systems that operate in the same band. Moreover, it was shown in \cite{Hoydis2013a} and \cite{Ngo2013a} that the transmit uplink/downlink powers can be reduced as $\frac{1}{\sqrt{N}}$ with only a minor loss in throughput. This allows for major reductions in the emitted power, but is actually bad from an overall energy efficiency (EE) perspective---the EE is maximized by increasing the emitted power with $N$ to compensate for the increasing circuit power consumption \cite{Bjornson2014b}.

This paper explores whether the huge degrees-of-freedom offered by massive MIMO provide robustness to transceiver hardware imperfections/impairments; for example, phase noise, non-linearities, quantization errors, noise amplification, and inter-carrier interference. Robustness to hardware imperfections has been conjectured in overview articles, such as \cite{Larsson2014a}. Such a characteristic is notably important since the deployment cost and circuit power consumption of massive MIMO scales linearly with $N$, unless the hardware accuracy constraints can be relaxed
such that low-power, low-cost hardware is deployed which is more prone to imperfections. Constant envelope precoding was analyzed in \cite{Mohammed2013a} to facilitate the use of power-efficient amplifiers in the downlink, while the impact of phase-drifts was analyzed and simulated for single-carrier systems in \cite{Pitarokoilis2014a} and for orthogonal frequency-division multiplexing (OFDM) in \cite{Krishnan2015a}. A preliminary proof of the conjecture was given in \cite{Bjornson2014a}, but the authors therein considered only additive distortions and, thus, ignored other important characteristics of hardware imperfections. That paper showed that one can tolerate distortion variances that increase as $\sqrt{N}$ with only minor throughput losses, but did not investigate what this implies for the design of different transceiver circuits.

\begin{figure}
\begin{center}
\includegraphics[width=.9\columnwidth]{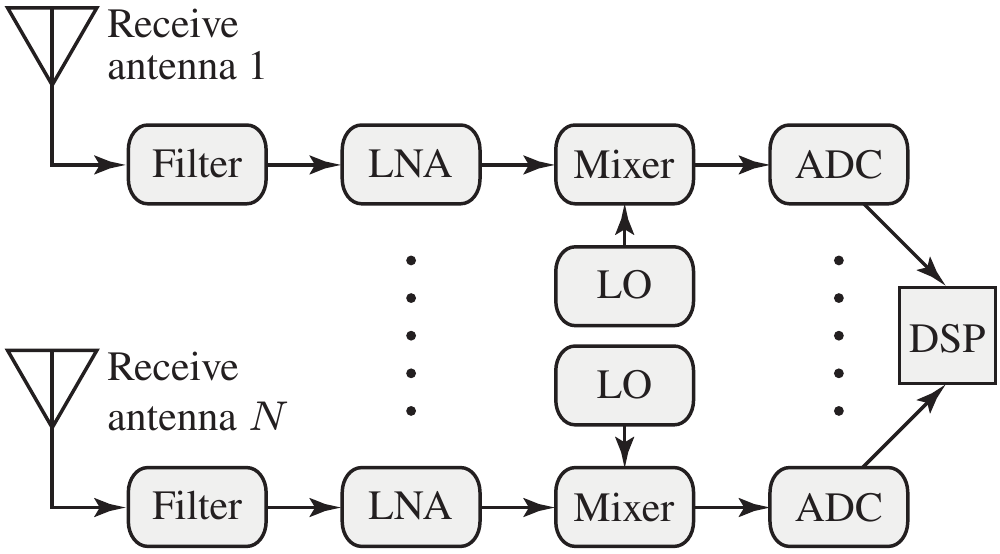}
\end{center}
\caption{Block diagram of a typical $N$-antenna receiver. The main circuits are shown, but these can be complemented with additional intermediate filters and amplifiers depending on the implementation. Most of the circuits affect only one antenna, whilst the LO can be either common for all antennas or different.} \label{figure_typicalreceiver}
\end{figure}

In this paper, we consider a generalized uplink massive MIMO system with arbitrary array configurations (e.g., co-located or distributed antennas). Based on the extensive literature on modeling of transceiver hardware imperfections (see \cite{Schenk2008a,Mezghani2010a,wenk2010mimo,Petrovic2007a,Bjornson2013d,Holma2011a,Mehrpouyan2012a,Durisi2013b,Pitarokoilis2014a,Zhang2012a} and references therein), we propose a tractable system model that jointly describes the impact of multiplicative phase-drifts, additive distortion noise, noise amplification, and inter-carrier interference. This stands in contrast to the previous works \cite{Pitarokoilis2014a,Bjornson2014a,Krishnan2015a}, which each investigated only one of these effects. The following are the main contributions of this paper:

\begin{itemize}
\item We derive a new linear minimum mean square error (LMMSE) channel estimator that accounts for hardware imperfections and allows the prediction of the detrimental impact of phase-drifts.

\item We present a simple and general expression for the achievable uplink user rates and compute it in closed-form, when the receiver applies maximum ratio combining (MRC) filters. We prove that the additive distortion noise and noise amplification vanish asymptotically as $N \rightarrow \infty$, while the phase-drifts remain but are not exacerbated.

\item We obtain an intuitive scaling law that shows how fast we can tolerate the levels of hardware imperfections to increase with $N$, while maintaining high user rates. This is an analytic proof of the conjecture that massive MIMO systems can be deployed with inexpensive low-power hardware without sacrificing the expected major performance gains. The scaling law provides sufficient conditions that hold for any judicious receive filters.

\item The practical implications of the scaling law are exemplified for the main circuits at the receiver, namely, the analog-to-digital converter (ADC), low noise amplifier (LNA), and local oscillator (LO). The main components of a typical receiver are illustrated in Fig.~\ref{figure_typicalreceiver}. The scaling law reveals the tradeoff between hardware cost, level of imperfections, and circuit power consumption. In particular, it shows how a circuit-aware design can make the circuit power consumption increase as $\sqrt{N}$ instead of $N$.

\item The analytic results are validated numerically in a realistic simulation setup, where we consider different antenna deployment scenarios, common and separate LOs, different pilot sequence designs, and two types of receive filters. A key observation is that separate LOs can provide better performance than a common LO, since the phase-drifts average out and the interference is reduced. This is also rigorously supported by the analytic scaling law.

\end{itemize}

This paper extends substantially our conference papers \cite{Bjornson2014c} and \cite{Bjornson2014d}, by generalizing the propagation model, generalizing the analysis according to the new model, and providing more comprehensive simulations.
The paper is organized as follows: In Section~\ref{sec:system-model}, the massive MIMO system model under consideration is presented. In Section~\ref{sec:performance-analysis}, a detailed performance analysis of the achievable uplink user rates is pursued and the impact of hardware imperfections is characterized, while
in Section~\ref{sec:circuit-examples} we provide guidelines for circuit-aware design in order to minimize the power
dissipation of receiver circuits. Our theoretical analysis is corroborated with simulations in Section~\ref{sec:numerical-results}, while Section \ref{sec:conclusion} concludes the paper.

\emph{Notation:} The following notation is used throughout the paper: Boldface (lower case) is used for column vectors, $\vect{x}$, and (upper case) for matrices, $\vect{X}$. Let $\vect{X}^{\Ttran}$, $\vect{X}^*$, and $\vect{X}^{\Htran}$ denote the transpose, conjugate, and conjugate transpose of $\vect{X}$, respectively. A diagonal matrix with $a_{1},\ldots,a_{N}$ on the main diagonal is denoted as $\diag(a_{1},\ldots,a_{N})$, while $\vect{I}_N$ is an $N \times N$ identity matrix. The set of complex-valued $N\times K$ matrices is denoted by $\mathbb{C}^{N\times K}$.
The expectation operator is denoted $\mathbb{E}\{\cdot\}$ and $\triangleq$ denotes definitions. The matrix trace function is $\tr(\cdot)$ and $\kron$ is the Kronecker product.
A Gaussian random variable $x$ is denoted $x \sim\mathcal{N}(\bar{x},q)$, where $\bar{x}$ is the mean and $q$ is the variance.
A circularly symmetric complex Gaussian random vector $\vect{x}$ is denoted $\vect{x} \sim \mathcal{CN}(\bar{\vect{x}},\vect{Q})$, where $\bar{\vect{x}}$ is the mean and $\vect{Q}$ is the covariance matrix. The big $\mathcal{O}$ notation $f(x) = \mathcal{O}(g(x))$ means that $\left| \frac{f(x)}{g(x)}\right|$ is bounded as $x\rightarrow \infty$.

\section{System Model with Hardware Imperfections} \label{sec:system-model}

We consider the uplink of a cellular network with $L\geq 1$ cells. Each cell consists of $K$ single-antenna user equipments (UEs) that communicate simultaneously with an array of $N$ antennas, which can be either co-located at a macro BS or distributed over multiple fully coordinated small BSs. The analysis of our paper holds for any $N$ and $K$, but we are primarily interested in massive MIMO topologies, where $N \gg K \gg 1$. The frequency-flat channel from UE $k$ in cell $l$ to BS $j$ is denoted as $\vect{h}_{jlk} \triangleq \left[h_{jlk}^{(1)} \, \ldots \, h_{jlk}^{(N)}\right]^{\Ttran} \in \mathbb{C}^{N \times 1}$ and is modeled as Rayleigh block fading. This means that it has a static realization for a coherence block of $T$ channel uses and independent realizations between blocks.\footnote{The size of the time/frequency block where the channels are static depends on UE mobility and propagation environment: $T$ is the product of the coherence time $\tilde{\tau}_c$ and coherence bandwidth $\tilde{W}_c$, thus $\tilde{\tau}_c = 5$ ms and $\tilde{W}_c = 100$ kHz gives $T=500$.} The UEs' channels are independent. Each realization is complex Gaussian distributed with zero mean and covariance matrix $\vect{\Lambda}_{jlk} \in \mathbb{C}^{N \times N}$:
\begin{equation}
\vect{h}_{jlk} \sim \mathcal{CN}(\vect{0}, \vect{\Lambda}_{jlk} ).
\end{equation}
The covariance matrix $\vect{\Lambda}_{jlk} \triangleq \diag\left( \lambda_{jlk}^{(1)},\ldots,\lambda_{jlk}^{(N)} \right)$ is assumed to be diagonal, which holds if the inter-antenna distances are sufficiently large and the multi-path scattering environment is rich \cite{Gao2015a}.\footnote{The analysis and main results of this paper can be easily extended to arbitrary non-diagonal covariance matrices as in \cite{Hoydis2013a} and \cite{Bjornson2014a}, but at the cost of complicating the notation and expressions.} The average channel attenuation $\lambda_{jlk}^{(n)}$ is different for each combination of cells, UE index, and receive antenna index $n$. It depends, for example, on the array geometry and the UE location. Even for co-located antennas one might have different values of $\lambda_{jlk}^{(n)}$ over the array, because of the large aperture that may create variations in the shadow fading.

The received signal $\vect{y}_j(t) \in \mathbb{C}^{N \times 1}$ in cell $j$ at a given channel use $t \in \{ 1,\ldots,T \}$ in the coherence block is conventionally modeled as \cite{Marzetta2010a,Rusek2013a,Hoydis2013a,Ngo2013a}
\begin{equation} \label{eq:conventional-model}
\vect{y}_j(t) = \sum_{l=1}^{L} \vect{H}_{jl} \vect{x}_{l}(t) + \vect{n}_{j}(t)
\end{equation}
where the transmit signal in cell $l$ is $\vect{x}_{l}(t) = [x_{l1}(t) \, \ldots \, x_{lK}(t)]^{\Ttran} \in \mathbb{C}^{K \times 1}$ and we use the notation $\vect{H}_{jl} = [ \vect{h}_{jl1} \, \ldots \, \vect{h}_{jlK}] \in \mathbb{C}^{N \times K}$ for brevity. The scalar signal $x_{lk}(t)$ sent by UE $k$ in cell $l$ at channel use $t$ is either a deterministic pilot symbol (used for channel estimation) or an information symbol from a Gaussian codebook; in any case, we assume that the expectation of the transmit energy per symbol is bounded as $\mathbb{E}\{ |x_{lk}(t) |^2 \} \leq p_{lk}$. The thermal noise vector $\vect{n}_{j}(t) \sim \mathcal{CN}(\vect{0},\sigma^2 \vect{I}_N)$ is spatially and temporally independent and has variance $\sigma^2$.

The conventional model in \eqref{eq:conventional-model} is well-accepted for small-scale MIMO systems, but has an important drawback when applied to massive MIMO topologies: it assumes that the large antenna array consists of $N$ high-quality antenna branches which are all perfectly synchronized. Consequently, the deployment cost and total power consumption of the circuits attached to each antenna would \emph{at least} grow linearly with $N$, thereby making the deployment of massive MIMO rather questionable, if not prohibitive, from an overall cost and efficiency perspective.

In this paper, we analyze the far more realistic scenario of having inexpensive hardware-constrained massive MIMO arrays. More precisely, each receive array experiences hardware imperfections that distort the communication. The exact distortion characteristics depend generally on which modulation scheme is used; for example, OFDM \cite{Schenk2008a}, filter bank multicarrier (FBMC) \cite{Farhang2014a}, or single-carrier transmission \cite{Pitarokoilis2014a}. Nevertheless, the distortions can be classified into three distinct categories: 1) received signals are shifted in phase; 2) distortion noise is added with a power proportional to the total received signal power; and 3) thermal noise is amplified and channel-independent interference is added. To draw general conclusions on how these distortion categories affect massive MIMO systems, we consider a generic system model with hardware imperfections. The received signal in cell $j$ at a given channel use $t \in \{ 1,\ldots,T \}$  is modeled as
\begin{equation} \label{eq:generalized-model}
\vect{y}_j(t) = \vect{D}_{\boldsymbol{\phi}_j(t)} \sum_{l=1}^{L} \vect{H}_{jl} \vect{x}_{l}(t) + \boldsymbol{\upsilon}_{j}(t) + \boldsymbol{\eta}_{j}(t)
\end{equation}
where the channel matrices $\vect{H}_{jl}$ and transmitted signals $\vect{x}_{l}(t)$ are exactly as in \eqref{eq:conventional-model}. The hardware imperfections are defined as follows:

\begin{enumerate}

\item The matrix $\vect{D}_{\boldsymbol{\phi}_j(t)} \! \triangleq \! \diag\left(e^{\imath \phi_{j1}(t)},\ldots, e^{\imath \phi_{jN}(t)}\right)$ describes multiplicative phase-drifts, where  $\imath$ is the  imaginary unit. The variable $\phi_{jn}(t)$ is the phase-drift at the $n$th receive antenna in cell $j$ at time $t$. Motivated by the standard phase-noise models in LOs \cite{Petrovic2007a}, $\phi_{jn}(t)$ follows a Wiener process
    \begin{equation}
    \phi_{jn}(t)  \sim  \mathcal{N}( \phi_{jn}(t - 1), \delta)
    \end{equation}
    which equals the previous realization $\phi_{jn}(t-1)$ plus an independent Gaussian innovation of variance $\delta$.
The phase-drifts can be either independent or correlated between the antennas; for example, co-located arrays might have a common LO (CLO) for all antennas which makes the phase-drifts $\phi_{jn}(t)$ identical for all $n=1,\ldots,N$. In contrast, distributed arrays might have separate LOs (SLOs) at each antenna, which make the drifts independent, though we let the variance $\delta$ be equal for simplicity. Both cases are considered herein.

\item The distortion noise $\boldsymbol{\upsilon}_{j}(t) \sim \mathcal{CN}(\vect{0},\vect{\Upsilon}_j(t) )$, where
\begin{equation}
\!\!\!\!\!\!\!\!\!\!\! \vect{\Upsilon}_j(t) \triangleq \kappa^2 \sum_{l=1}^{L} \sum_{k=1}^{K} \mathbb{E}\{ |x_{lk}(t) |^2 \} \diag\bigg( |h_{jlk}^{(1)}|^2,\ldots, |h_{jlk}^{(N)}|^2\bigg)
\end{equation}
for given channel realizations, where the double-sum gives the received power at each antenna. Thus, the distortion noise is independent between antennas and channel uses, and the variance at a given antenna is proportional to the current received signal power at this antenna. This model can describe the quantization noise in ADCs with gain control \cite{Mezghani2010a}, approximate generic non-linearities \cite[Chapter 14]{Holma2011a}, and approximate the leakage between subcarriers due to calibration errors. The parameter $\kappa \geq 0$ describes how much weaker the distortion noise magnitude is compared to the signal magnitude.

\item The receiver noise $\boldsymbol{\eta}_{j}(t) \sim \mathcal{CN}(\vect{0},\xi \vect{I}_N)$ is independent of the UE channels, in contrast to the distortion noise. This term includes thermal noise, which typically is amplified by LNAs and mixers in the receiver hardware, and interference leakage from other frequency bands and/or other networks. The receiver noise variance must satisfy $\xi \geq \sigma^2$. If there is no interference leakage, $F=\frac{\xi}{\sigma^2}$ is called the noise amplification factor.
\end{enumerate}

This tractable generic model of hardware imperfections at the BSs is inspired by a plethora of prior works \cite{Schenk2008a,Mezghani2010a,wenk2010mimo,Petrovic2007a,Bjornson2013d,Holma2011a,Mehrpouyan2012a,Durisi2013b,Pitarokoilis2014a,Zhang2012a} and characterizes the joint behavior of all hardware imperfections at the BSs---these can be uncalibrated imperfections or residual errors after calibration. The model in \eqref{eq:generalized-model} is characterized by three parameters: $\delta$, $\kappa$, and $\xi$. The model is compatible with the conventional model in \eqref{eq:conventional-model}, which is obtained by setting $\xi = \sigma^2$ and $\delta \!=\! \kappa \!=\! 0$. The analysis in this paper holds for arbitrary parameter values. Section \ref{sec:circuit-examples} exemplifies the connection between imperfections in the main transceiver circuits of the BSs and the three parameters. These connections allow for circuit-aware design of massive MIMO systems.

In the next section, we derive a channel estimator and achievable UE rates for the system model in \eqref{eq:generalized-model}. By analyzing the performance as $N\rightarrow\infty$, we bring new insights into the fundamental impact of hardware imperfections (in particular, in terms of $\delta$, $\kappa$, and $\xi$).

\section{Performance Analysis}
\label{sec:performance-analysis}

\begin{figure}
\begin{center}
\includegraphics[width=\columnwidth]{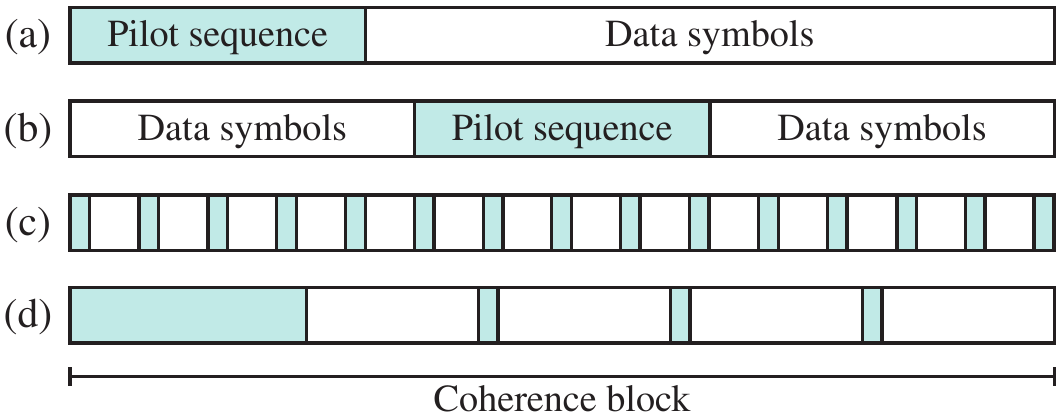}
\end{center}
\caption{Examples of different ways to distribute the $B$ pilot symbols over the coherence block of length $T$: (a) beginning of block; (b) middle of block; (c) uniform pilot distribution; (d) preamble and a few distributed pilot symbols.} \label{figure_protocol}
\end{figure}

In this section, we derive achievable UE rates for the uplink multi-cell system in \eqref{eq:generalized-model} and analyze how these depend on the number of antennas and hardware imperfections. We first need to specify the transmission protocol.\footnote{We assume that the same protocol is used in all cells, for analytic simplicity. It was shown in \cite[Remark 5]{Ngo2013a} that nothing substantially different will happen if this assumption is relaxed.} The $T$ channel uses of each coherence block are split between transmission of uplink pilot symbols and uplink data symbols. It is necessary to dedicate $B \geq K$ channel uses for pilot transmission if the receiving array should be able to spatially separate the different UEs in the cell. The remaining $T-B$ channel uses are allocated for data transmission. The pilot symbols can be distributed in different ways: for example, placed in the beginning of the block \cite{Bjornson2014a}, in the middle of the block \cite{Yang2013a}, uniformly distributed as in the LTE standard \cite{Dahlman2008a}, or a combination of these approaches \cite{Mehrpouyan2012a}. These different cases are illustrated in Fig.~\ref{figure_protocol}. The time indices used for pilot transmission are denoted by $\tau_1,\ldots,\tau_B \in \{1,\ldots,T\}$, while $\mathcal{D} \triangleq \{1,\ldots,T\} \setminus \{ \tau_1,\ldots,\tau_B \}$ are the time indices for data transmission.

\subsection{Channel Estimation under Hardware Imperfections}

Based on the transmission protocol, the pilot sequence of UE $k$ in cell $j$ is $\tilde{\vect{x}}_{jk} \triangleq [x_{jk}(\tau_1) \, \ldots \, x_{jk}(\tau_B)]^{\Ttran} \in \mathbb{C}^{B \times 1}$. The pilot sequences are predefined and can be selected arbitrarily under the power constraints. Our analysis supports any choice, but it is reasonable to make  $\tilde{\vect{x}}_{j1},\ldots,\tilde{\vect{x}}_{jK}$ in cell $j$ mutually orthogonal to avoid intra-cell interference (this is the reason to have $B \geq K$).

\begin{example} \label{example:pilot-sequences}
Let $\widetilde{\vect{X}}_j \triangleq [\tilde{\vect{x}}_{j1} \, \ldots \, \tilde{\vect{x}}_{jK}] $ denote the pilot sequences in cell $j$. The simplest example of linearly independent pilot sequences (with $B=K$) is
\begin{equation} \label{eq:temporal-orthogonal-matrix}
\widetilde{\vect{X}}_j^{\mathrm{temporal}} \triangleq \diag(\sqrt{p_{j1}},\ldots,\sqrt{p_{jK}})
\end{equation}
where the different sequences are temporally orthogonal since only UE $k$ transmits at time $\tau_k$. Alternatively, the pilot sequences can be made spatially orthogonal so that all UEs transmit at every pilot transmission time, which effectively increases the total pilot energy by a factor $K$. The canonical example is to use a scaled discrete Fourier transform (DFT) matrix \cite{Biguesh2004a}:
\begin{equation} \label{eq:dft-pilot-matrix}
\widetilde{\vect{X}}_j^{\mathrm{spatial}} \triangleq \begin{bmatrix}
1 & 1 & \!\!\ldots\!\! &1 \\
1 & W_K & \!\!\ldots\!\! & W_K^{K-1} \\
\vdots & \!\!\vdots\!\! & \vdots & \vdots \\
1 & W_K^{B-1} & \!\!\ldots\!\! & W_K^{(B-1)(K-1)}
      \end{bmatrix} \widetilde{\vect{X}}_j^{\mathrm{temporal}}
\end{equation}
where $W_K \triangleq e^{-\imath 2 \pi / K}$.
\end{example}

\begin{figure*}[t!]
\begin{align} \label{eq:achievable-SINR}
\mathrm{SINR}_{jk}(t) =  \frac{ p_{jk} | \mathbb{E}\{ \vect{v}_{jk}^{\Htran}(t) \vect{h}_{jjk}(t) \} |^2 }{ \fracSumtwo{l=1}{L} \fracSumtwo{m=1}{K} p_{lm}  \mathbb{E}\{ |\vect{v}_{jk}^{\Htran}(t) \vect{h}_{jlm}(t) |^2  \} - p_{jk} | \mathbb{E}\{ \vect{v}_{jk}^{\Htran}(t) \vect{h}_{jjk}(t) \} |^2 + \mathbb{E}\{ |\vect{v}_{jk}^{\Htran}(t) \boldsymbol{\upsilon}_j(t) |^2  \}   +  \xi \mathbb{E}\{ \| \vect{v}_{jk}(t) \|^2\}  } \tag{20}
\end{align}
\hrulefill
\end{figure*}

The pilot sequences can also be jointly designed across cells, to reduce inter-cell interference during pilot transmission. Since network-wide pilot orthogonality requires $B \geq LK$, which typically is much larger than the coherence block length $T$, practical networks need to balance between pilot orthogonality and inter-cell interference. A key design goal is to allocate non-orthogonal pilot sequences to UEs that have nearly orthogonal channel covariance matrices; for example, by making $\tr(\vect{\Lambda}_{jjk} \vect{\Lambda}_{jlm})$ small for any combination of a UE $k$ in cell $j$ and a UE $m$ in cell $l$, as suggested in \cite{Yin2013a}.

For any given set of pilot sequences, we now derive estimators of the effective channels
\begin{equation}
\vect{h}_{jlk}(t) \triangleq \vect{D}_{\boldsymbol{\phi}_j(t)} \vect{h}_{jlk}
\end{equation}
at any channel use $t \in \{1,\ldots,T\}$ and for all $j,l,k$. The conventional multi-antenna channel estimators from \cite{Kay1993a,Kotecha2004a,Bjornson2010a} cannot be applied in this paper since the generalized system model in \eqref{eq:generalized-model} has two non-standard properties: the pilot transmission is corrupted by random phase-drifts and the distortion noise is statistically dependent on the channels. Therefore, we derive a new LMMSE estimator for the system model at hand.

\begin{theorem} \label{theorem:LMMSE-estimation}
Let $\boldsymbol{\psi}_j \triangleq \left[\vect{y}_j^{\Ttran}(\tau_1) \, \ldots \, \vect{y}_j^{\Ttran}(\tau_B)\right]^{\Ttran} \in \mathbb{C}^{BN}$ denote the combined received signal in cell $j$ from the pilot transmission. The LMMSE estimate of $\vect{h}_{jlk}(t)$ at any channel use $t \in \{ 1, \ldots, T \}$ for any $l$ and $k$ is
\begin{equation} \label{eq:LMMSE-estimator}
  \hat{\vect{h}}_{jlk}(t) = \left( \tilde{\vect{x}}_{lk}^{\Htran}  \vect{D}_{\boldsymbol{\delta}(t)} \kron \vect{\Lambda}_{jlk} \right) \boldsymbol{\Psi}^{-1}_j \boldsymbol{\psi}_j
\end{equation}
where
\begin{align}
\vect{D}_{\boldsymbol{\delta}(t)} & \triangleq \diag\bigg( e^{-\frac{\delta}{2} |t-\tau_1|}, \ldots, e^{-\frac{\delta}{2} |t-\tau_B|}\bigg), \\
\boldsymbol{\Psi}_j &\triangleq \sum_{\ell=1}^{L} \sum_{m=1}^{K} \vect{X}_{\ell m} \kron \vect{\Lambda}_{j \ell m}  + \xi \vect{I}_{BN}, \\
\vect{X}_{\ell m} &\triangleq \bar{\vect{X}}_{\ell m} + \kappa^2 \vect{D}_{| \tilde{\vect{x}}_{\ell m}|^2}, \\
\vect{D}_{| \tilde{\vect{x}}_{\ell m}|^2} &\triangleq \diag\bigg( |x_{\ell m}(\tau_1)|^2 , \ldots, |x_{\ell m}(\tau_B)|^2 \bigg),
\end{align}
while the element $(b_1,b_2)$ of $\bar{\vect{X}}_{\ell m} \in \mathbb{C}^{B \times B}$ is
\begin{equation}
[\bar{\vect{X}}_{\ell m} ]_{b_1,b_2} = \begin{cases} |x_{\ell m}(\tau_{b_1})|^2, & b_1 = b_2, \\ x_{\ell m}(\tau_{b_1}) x_{\ell m}^*(\tau_{b_2}) e^{-\frac{\delta}{2} |\tau_{b_1}-\tau_{b_2}|}, &  b_1 \neq b_2. \end{cases}
\end{equation}
The corresponding error covariance matrix is
\begin{align} 
\vect{C}_{jlk}(t) &= \mathbb{E}\left\{ \big( \vect{h}_{jlk}(t) -\hat{\vect{h}}_{jlk}(t) \big) \big( \vect{h}_{jlk}(t) - \hat{\vect{h}}_{jlk}(t) \big)^{\Htran}   \right\} \notag \\ &=
\vect{\Lambda}_{jlk} - \left( \tilde{\vect{x}}_{lk}^{\Htran}  \vect{D}_{\boldsymbol{\delta}(t)} \kron \vect{\Lambda}_{jlk} \right) \boldsymbol{\Psi}^{-1}_j
(  \vect{D}_{\boldsymbol{\delta}(t)}^{\Htran} \tilde{\vect{x}}_{lk}   \kron \vect{\Lambda}_{jlk} ) \label{eq:LMMSE-error-cov}
\end{align}
and the mean-squared error (MSE) is $\mathrm{MSE}_{jlk}(t) = \tr( \vect{C}_{jlk}(t) )$.
\end{theorem}
\begin{IEEEproof}
The proof is given in Appendix B.
\end{IEEEproof}

It is important to note that although the channels are block fading, the phase-drifts caused by hardware imperfections make the effective channels $\vect{h}_{jlk}(t)$ change between every channel use. The new LMMSE estimator in Theorem \ref{theorem:LMMSE-estimation} provides different estimates for each time index $t \in \mathcal{D}$ used for data transmission---this is a prediction, interpolation, or retrospection depending on how the pilot symbols are distributed in the coherence block (recall Fig.~\ref{figure_protocol}). The LMMSE estimator is the same for systems with independent and correlated phase-drifts which brings robustness to modeling errors, but also means that there exist better non-linear estimators that can exploit phase-drift correlations, though we do not pursue this issue further in this paper.

The estimator expression is simplified in the special case of co-located arrays, as shown by the following corollary.

\begin{corollary}
If  $\vect{\Lambda}_{jlk} = \lambda_{jlk} \vect{I}_N$ for all $j$, $l$, and $k$, the LMMSE estimate in \eqref{eq:LMMSE-estimator} simplifies to
\begin{equation} \label{eq:LMMSE-estimator-simplified}
  \hat{\vect{h}}_{jlk}(t) = \bigg( \lambda_{jlk} \tilde{\vect{x}}_{lk}^{\Htran}  \vect{D}_{\boldsymbol{\delta}(t)} \boldsymbol{\Omega}^{-1}_j \kron \vect{I}_N \bigg) \boldsymbol{\psi}_j
\end{equation}
and the error covariance matrix in \eqref{eq:LMMSE-error-cov} becomes
\begin{equation} \label{eq:LMMSE-error-cov-simplified}
\begin{split}
\vect{C}_{jlk}(t) = \lambda_{jlk} \bigg( 1 - \lambda_{jlk} \tilde{\vect{x}}_{lk}^{\Htran}  \vect{D}_{\boldsymbol{\delta}(t)} \boldsymbol{\Omega}^{-1}_j \vect{D}_{\boldsymbol{\delta}(t)}^{\Htran} \tilde{\vect{x}}_{lk} \bigg) \vect{I}_N
\end{split}
\end{equation}
where $\boldsymbol{\Omega}_j$ is the Hermitian matrix
\begin{equation}
\boldsymbol{\Omega}_j \triangleq \sum_{\ell=1}^{L} \sum_{m=1}^{K} \lambda_{j \ell m} \vect{X}_{\ell m} + \xi \vect{I}_B.
\end{equation}
\end{corollary}

Next, we use these channel estimates to design receive filters and derive achievable UE rates.

\subsection{Achievable UE Rates under Hardware Imperfections}
\label{subsec:user-rates}

It is difficult to compute the maximum achievable UE rates when the receiver has imperfect channel knowledge \cite{Yoo2006b}, and hardware imperfections are not simplifying this task. Upper bounds on the achievable rates were obtained in \cite{Bjornson2014a} and \cite{Durisi2014a}. In this paper, we want to guarantee certain performance and thus seek simple achievable (but suboptimal) rates. The following lemma provides such rate expressions and builds upon well-known techniques from \cite{Medard2000a,Hassibi2003a,Yoo2006b,Marzetta2010a,Pitarokoilis2014a} for computing lower bounds on the mutual information.

\begin{lemma} \label{lemma:achievable-rates}
Suppose the receiver in cell $j$ has complete statistical channel knowledge and applies the linear receive filters $\vect{v}_{jk}^{\Htran}(t) \in \mathbb{C}^{1 \times N}$, for $t \in \mathcal{D}$, to detect the signal from its $k$th UE. An ergodic achievable rate for this UE is
\begin{equation} \label{eq:achievable-rate}
R_{jk} = \frac{1}{T}  \sum_{t \in \mathcal{D}} \log_2 \big( 1 + \mathrm{SINR}_{jk}(t) \big) \quad [\textrm{bit/channel use}]
\end{equation}
where $\mathrm{SINR}_{jk}(t)$ is given in \eqref{eq:achievable-SINR} at the top of this page and all UEs use full power (i.e., $\mathbb{E}\{ |x_{lk}(t) |^2 \} = p_{lk}$ for all $l,k$).
\end{lemma}
\begin{IEEEproof}
The proof is given in Appendix C.
\end{IEEEproof}

\setcounter{equation}{20}

The achievable UE rates in Lemma \ref{lemma:achievable-rates} can be computed for any choice of receive filters, using numerical methods; the MMSE receive filter is simulated in Section \ref{sec:numerical-results}.
Note that the sum in \eqref{eq:achievable-rate} has $|\mathcal{D}| = T-B$ terms, while the pre-log factor $\frac{1}{T}$ also accounts for the $B$ channel uses of pilot transmissions. The next theorem gives new closed-form expressions for all the expectations in \eqref{eq:achievable-SINR} when using MRC receive filters.

\begin{figure*}[t!]
\begin{align}
\mathbb{E}\{ \|\vect{v}_{jk}(t) \|^2\} & = \tr \left( \left( \tilde{\vect{x}}_{jk}^{\Htran}  \vect{D}_{\boldsymbol{\delta}(t)} \kron \vect{\Lambda}_{jjk} \right) \boldsymbol{\Psi}^{-1}_j \left(  \vect{D}_{\boldsymbol{\delta}(t)}^{\Htran} \tilde{\vect{x}}_{jk}   \kron \vect{\Lambda}_{jjk} \right) \right)  \label{eq:MRC-squared-norm} \\
\mathbb{E}\{ \vect{v}_{jk}^{\Htran}(t) \vect{h}_{jjk}(t) \} & = \mathbb{E}\{ \|\vect{v}_{jk}(t) \|^2\}  \label{eq:MRC-first-moment} \\
\mathbb{E}\{ | \vect{v}_{jk}^{\Htran}(t) \vect{h}_{jlm}(t) |^2 \} &= \tr \left( \vect{\Lambda}_{jlm} \left( \tilde{\vect{x}}_{jk}^{\Htran}  \vect{D}_{\boldsymbol{\delta}(t)} \kron \vect{\Lambda}_{jjk} \right) \boldsymbol{\Psi}^{-1}_j \left(  \vect{D}_{\boldsymbol{\delta}(t)}^{\Htran} \tilde{\vect{x}}_{jk}   \kron \vect{\Lambda}_{jjk} \right) \right) \label{eq:MRC-second-moment} \\
&\!\!\!\!\!\!\!\!\!\!\!\!\!\!\!\!\!\!\!\!\!\!\!\!\!\!\!\!\!\!\!\!\!\!\!\!\!\!\!\! + \begin{cases}
\fracSumtwo{n_1=1}{N} \fracSumtwo{n_2=1}{N} \lambda_{jjk}^{(n_1)} \lambda_{jlm}^{(n_1)} \lambda_{jjk}^{(n_2)}  \lambda_{jlm}^{(n_2)}
 \left( \tilde{\vect{x}}_{jk}^{\Htran}  \vect{D}_{\boldsymbol{\delta}(t)} \kron \vect{e}_{n_1}^{\Htran} \right) \boldsymbol{\Psi}^{-1}_j \left( \bar{\vect{X}}_{lm} \kron \vect{e}_{n_1} \vect{e}_{n_2}^{\Htran} \right) \boldsymbol{\Psi}^{-1}_j \left(  \vect{D}_{\boldsymbol{\delta}(t)}^{\Htran}  \tilde{\vect{x}}_{jk} \kron \vect{e}_{n_2} \right) & \text{if a CLO}\\
\left( \tr \left( \left( \tilde{\vect{x}}_{jk}^{\Htran}  \vect{D}_{\boldsymbol{\delta}(t)} \kron \vect{\Lambda}_{jjk} \right) \boldsymbol{\Psi}^{-1}_j \left(  \vect{D}_{\boldsymbol{\delta}(t)}^{\Htran} \tilde{\vect{x}}_{lm}   \kron \vect{\Lambda}_{jlm} \right) \right)   \right)^2 & \text{if SLOs}
\end{cases} \notag \\
&\!\!\!\!\!\!\!\!\!\!\!\!\!\!\!\!\!\!\!\!\!\!\!\!\!\!\!\!\!\!\!\!\!\!\!\!\!\!\!\! + \begin{cases}
\fracSumtwo{n=1}{N}  \left( \lambda_{jjk}^{(n)} \lambda_{jlm}^{(n)}  \right)^2
\left( \tilde{\vect{x}}_{jk}^{\Htran}  \vect{D}_{\boldsymbol{\delta}(t)} \kron \vect{e}_n^{\Htran} \right) \boldsymbol{\Psi}^{-1}_j
\left( \kappa^2 \vect{D}_{| \tilde{\vect{x}}_{l m}|^2} \kron \vect{e}_n \vect{e}_n^{\Htran} \right) \boldsymbol{\Psi}^{-1}_j \left(  \vect{D}_{\boldsymbol{\delta}(t)}^{\Htran}  \tilde{\vect{x}}_{jk} \kron \vect{e}_n \right) & \text{if a CLO}\\
\fracSumtwo{n=1}{N}  \left( \lambda_{jjk}^{(n)} \lambda_{jlm}^{(n)}  \right)^2
\left( \tilde{\vect{x}}_{jk}^{\Htran}  \vect{D}_{\boldsymbol{\delta}(t)} \kron \vect{e}_n^{\Htran} \right) \boldsymbol{\Psi}^{-1}_j
\left( (\vect{X}_{l m} - \vect{D}_{\boldsymbol{\delta}(t)}^{\Htran} \tilde{\vect{x}}_{lm}   \tilde{\vect{x}}_{lm}^{\Htran} \vect{D}_{\boldsymbol{\delta}(t)}) \kron \vect{e}_n \vect{e}_n^{\Htran} \right) \boldsymbol{\Psi}^{-1}_j \left(  \vect{D}_{\boldsymbol{\delta}(t)}^{\Htran}  \tilde{\vect{x}}_{jk} \kron \vect{e}_n \right)
& \text{if SLOs}
\end{cases} \notag \\
\mathbb{E}\{ |\vect{v}_{jk}^{\Htran}(t) \boldsymbol{\upsilon}_j(t) |^2  \}&= \kappa^2 \sum_{l=1}^{L} \sum_{m=1}^{K} p_{lm} \tr \left( \vect{\Lambda}_{jlm} \left( \tilde{\vect{x}}_{jk}^{\Htran}  \vect{D}_{\boldsymbol{\delta}(t)} \kron \vect{\Lambda}_{jjk} \right) \boldsymbol{\Psi}^{-1}_j \left(  \vect{D}_{\boldsymbol{\delta}(t)}^{\Htran} \tilde{\vect{x}}_{jk}   \kron \vect{\Lambda}_{jjk} \right) \right) \label{eq:MRC-cross-moment}  \\
&+ \kappa^2 \sum_{l=1}^{L} \sum_{m=1}^{K}
\sum_{n=1}^{N} p_{lm}  \left( \lambda_{jjk}^{(n)} \lambda_{jlm}^{(n)} \right)^2
\left( \tilde{\vect{x}}_{jk}^{\Htran}  \vect{D}_{\boldsymbol{\delta}(t)} \kron \vect{e}_n^{\Htran} \right) \boldsymbol{\Psi}^{-1}_j
( \vect{X}_{l m}  \kron \vect{e}_n \vect{e}_n^{\Htran} ) \boldsymbol{\Psi}^{-1}_j \left(  \vect{D}_{\boldsymbol{\delta}(t)}^{\Htran}  \tilde{\vect{x}}_{jk} \kron \vect{e}_n \right) \notag
\end{align}
\hrulefill
\end{figure*}

\begin{theorem} \label{theorem:MRC-expectations}
The expectations in the SINR expression \eqref{eq:achievable-SINR} are given in closed form by \eqref{eq:MRC-squared-norm}--\eqref{eq:MRC-cross-moment}, at the top of the next page, when the MRC receive filter $\vect{v}^{\mathrm{MRC}}_{jk}(t) = \hat{\vect{h}}_{jjk}(t)$ is used in cell $j$. The $n$th column of $\vect{I}_N$ is denoted by $\vect{e}_n \in \mathbb{C}^{N \times 1}$ in this paper.
\end{theorem}
\begin{IEEEproof}
The proof is given in Appendix D.
\end{IEEEproof}

By substituting the expressions from Theorem \ref{theorem:MRC-expectations} into \eqref{eq:achievable-SINR}, we obtain closed-form UE rates that are achievable using MRC filters. Although the expressions in \eqref{eq:MRC-squared-norm}--\eqref{eq:MRC-cross-moment} are easy to compute, their interpretation is non-trivial. The size of each term depends on the setup and scales differently with $N$; note that each trace-expression and/or sum over the antennas give a scaling factor of $N$. This property is easily observed in the special case of co-located antennas:

\begin{corollary} \label{corollary:MRC-expectations-collocated}
If $\vect{\Lambda}_{jlk} = \lambda_{jlk} \vect{I}_N$ for all $j$, $l$, and $k$, the MRC receive filter yields:

\begin{align}
\mathbb{E}\{ \|\vect{v}_{jk}(t) \|^2\} & = N \lambda_{jjk}^2 \tilde{\vect{x}}_{jk}^{\Htran}  \vect{D}_{\boldsymbol{\delta}(t)} \boldsymbol{\Omega}^{-1}_j \vect{D}_{\boldsymbol{\delta}(t)}^{\Htran} \tilde{\vect{x}}_{jk} \notag \\
\mathbb{E}\{ \vect{v}_{jk}^{\Htran}(t) \vect{h}_{jjk}(t) \} & = \mathbb{E}\{ \|\vect{v}_{jk}(t) \|^2\}  \notag \\
\mathbb{E}\{ | \vect{v}_{jk}^{\Htran}(t) \vect{h}_{jlm}(t) |^2 \} &= \lambda_{jlm} \mathbb{E}\{ \| \vect{v}_{jk}(t) \|^2\}  \notag  \\
&\!\!\!\!\!\!\!\!\!\!\!\!\!\!\!\!\!\!\!\!\!\!\!\!\!\!\!\!\!\!\!\!\!\!\!\!\!\!\!\!\!\!\!\!\!\!\!\!\!\!\!\!  + N  \lambda_{jjk}^2 \lambda_{jlm}^2 \tilde{\vect{x}}_{jk}^{\Htran}   \vect{D}_{\boldsymbol{\delta}(t)} \boldsymbol{\Omega}_j^{-1} \vect{X}_{lm}  \boldsymbol{\Omega}_j^{-1} \vect{D}_{\boldsymbol{\delta}(t)}^{\Htran} \tilde{\vect{x}}_{jk} + N(N\!-\!1) \notag \\
&\!\!\!\!\!\!\!\!\!\!\!\!\!\!\!\!\!\!\!\!\!\!\!\!\!\!\!\!\!\!\!\!\!\!\!\!\!\!\!\!\!\!\!\!\!\!\!\!\!\!\!\!
\times\begin{cases}
\lambda_{jjk}^2 \lambda_{jlm}^2 \tilde{\vect{x}}_{jk}^{\Htran}   \vect{D}_{\boldsymbol{\delta}(t)} \boldsymbol{\Omega}_j^{-1} \bar{\vect{X}}_{lm}  \boldsymbol{\Omega}_j^{-1} \vect{D}_{\boldsymbol{\delta}(t)}^{\Htran} \tilde{\vect{x}}_{jk}  & \!\! \textrm{if a CLO} \\
 \lambda_{jjk}^2 \lambda_{jlm}^2 |\tilde{\vect{x}}_{jk}^{\Htran}  \vect{D}_{\boldsymbol{\delta}(t)} \boldsymbol{\Omega}_j^{-1}  \vect{D}_{\boldsymbol{\delta}(t)}^{\Htran} \tilde{\vect{x}}_{lm} |^2 & \!\! \textrm{if SLOs}
\end{cases} \notag \\
\mathbb{E}\{ |\vect{v}_{jk}^{\Htran}(t) \boldsymbol{\upsilon}_j(t) |^2  \}&= \kappa^2 \mathbb{E}\{ \|\vect{v}_{jk}(t) \|^2\} \sum_{l=1}^{L} \sum_{m=1}^{K} p_{lm} \lambda_{jlm}  \notag \\
&\!\!\!\!\!\!\!\!\!\!\!\!\!\!\!\!\!\!\!\!\!\!\!\!\!\!\!\!\!\!\!\!\!\!\!\!\!\!\!\!\!\!\!\!\!\!\!\!\!\!\!\!\! +\!\! \kappa^2 \!\sum_{l=1}^{L} \sum_{m=1}^{K} p_{lm} N  \lambda_{jjk}^2 \lambda_{jlm}^2 \tilde{\vect{x}}_{jk}^{\Htran}   \vect{D}_{\boldsymbol{\delta}(t)} \boldsymbol{\Omega}_j^{-1} \vect{X}_{lm}  \boldsymbol{\Omega}_j^{-1} \vect{D}_{\boldsymbol{\delta}(t)}^{\Htran} \tilde{\vect{x}}_{jk}. \notag
\end{align}
\end{corollary}

As seen from this corollary, most terms scale linearly with $N$ but there are a few terms that scale as $N^2$. The latter terms dominate in the asymptotic analysis below.

The difference between having a CLO and SLOs only manifests itself in the second-order moments $\mathbb{E}\{ | \vect{v}_{jk}^{\Htran}(t) \vect{h}_{jlm}(t) |^2 \}$. Hence, the desired signal quality is the same in both cases, while the interference terms are different; the case with the smallest interference variance $\sum_{l=1}^{L} \sum_{m=1}^{K} p_{lm}  \mathbb{E}\{ |\vect{v}_{jk}^{\Htran}(t) \vect{h}_{jlm}(t) |^2  \}$ gives the largest rate for UE $k$ in cell $j$. These second-order moments depend on the pilot sequences, channel covariance matrices, and phase-drifts. By looking at \eqref{eq:MRC-second-moment} in Theorem \ref{theorem:MRC-expectations} (or the corresponding expression in Corollary \ref{corollary:MRC-expectations-collocated}), we see that the only difference is that two occurrences of $\bar{\vect{X}}_{\ell m}$ in the case of a CLO are replaced by $\vect{D}_{\boldsymbol{\delta}(t)}^{\Htran}  \tilde{\vect{x}}_{\ell m} \tilde{\vect{x}}_{\ell m}^{\Htran} \vect{D}_{\boldsymbol{\delta}(t)} $ in the case of SLOs. These terms are equal when there are no phase-drifts (i.e., $\delta=0$), while the difference grows larger with $\delta$.
In particular, the term $\bar{\vect{X}}_{\ell m}$ is unaffected by the time index $t$, while the corresponding terms for SLOs decay as $e^{-\delta t}$ (from $\vect{D}_{\boldsymbol{\delta}(t)}$). The following example provides the intuition behind this result.

\begin{example} \label{example:CLOvsSLOs}
The interference power in \eqref{eq:achievable-SINR} consists of multiple terms of the form $\mathbb{E}\{ |\vect{v}_{jk}^{\Htran}(t) \vect{D}_{\boldsymbol{\phi}_j(t)} \vect{h}_{jlm} |^2  \}$. Suppose that the receive filter is set to some constant $\vect{v}_{jk}(t) = \vect{v}_{jk}$. If a CLO is used, we have $\mathbb{E}\{ |\vect{v}_{jk}^{\Htran} \vect{D}_{\boldsymbol{\phi}_j(t)} \vect{h}_{jlm} |^2  \} = \mathbb{E}\{ |\vect{v}_{jk}^{\Htran} \vect{h}_{jlm} |^2  \} $, which is independent of the phase-drifts since all elements of $\vect{v}_{jk}$ are rotated in the same way. In contrast, each component of $\vect{v}_{jk}$ is rotated in an independent random manner with SLOs, which reduces the average interference power since the components of the inner product $\vect{v}_{jk}^{\Htran} \vect{D}_{\boldsymbol{\phi}_j(t)} \vect{h}_{jlm}$ add up incoherently. Consequently, the received interference power is reduced by SLOs while it remains the same with a CLO.
\end{example}

To summarize, we expect SLOs to provide larger UE rates than a CLO, because the interference reduces with $t$ when the phase-drifts are independent, at the expense of increasing the deployment cost by having $N$ LOs. This observation is validated by simulations in Section \ref{sec:numerical-results}.

\subsection{Asymptotic Analysis and Hardware Scaling Laws}

The closed-form expressions in Theorem \ref{theorem:MRC-expectations} and Corollary \ref{corollary:MRC-expectations-collocated} can be applied to cellular networks of arbitrary (finite) dimensions. In massive MIMO, the asymptotic behavior of large antenna arrays is of particular interest. In this section, we assume that the $N$ receive antennas in each cell are distributed over $A\geq 1$ spatially separated subarrays, where each subarray contains $\frac{N}{A}$ antennas. This assumption is made for analytic tractability, but also makes sense in many practical scenarios. Each subarray is assumed to have an inter-antenna distance much smaller than the propagation distances to the UEs, such that $ \tilde{\lambda}_{jlk}^{(a)}$ is the average channel attenuation to all antennas in subarray $a$ in cell $j$ from UE $k$ in cell $l$.
Hence, the channel covariance matrix $\vect{\Lambda}_{jlk} \in \mathbb{C}^{N \times N}$ can be factorized as
\begin{equation} \label{eq:covariance-matrices-asymptotics}
\vect{\Lambda}_{jlk}
      = \underbrace{\diag\bigg( \tilde{\lambda}_{jlk}^{(1)},\ldots,\tilde{\lambda}_{jlk}^{(A)}\bigg)}_{\triangleq \tilde{\vect{\Lambda}}_{jlk}^{(A)}  \in \mathbb{C}^{A \times A} }  \kron \vect{I}_{\frac{N}{A}}.
\end{equation}
By letting the number of antennas in each subarray grow large, we obtain the following property.

\begin{corollary} \label{corollary:asymptotic-SINR}
If the MRC receive filter is used and the channel covariance matrices can be factorized as in \eqref{eq:covariance-matrices-asymptotics}, then
\begin{equation} \label{eq:asymptotic-SINR}
\mathrm{SINR}_{jk}(t) = \frac{ p_{jk} \mathrm{Sig}_{jk} }{\fracSumtwo{l=1}{L} \fracSumtwo{m=1}{K} p_{lm} \mathrm{Int}_{jklm}   - p_{jk} \mathrm{Sig}_{jk} \!+\! \mathcal{O}\left(\frac{1}{N}\right) }
\end{equation}
where the signal part is
\begin{equation}
\mathrm{Sig}_{jk} = \left( \tr \left( \left( \tilde{\vect{x}}_{jk}^{\Htran}  \vect{D}_{\boldsymbol{\delta}(t)} \!\kron\! \tilde{\vect{\Lambda}}_{jjk}^{(A)} \right) \widetilde{\boldsymbol{\Psi}}^{-1}_j \left(  \vect{D}_{\boldsymbol{\delta}(t)} \tilde{\vect{x}}_{jk}  \!\kron\! \tilde{\vect{\Lambda}}_{jjk}^{(A)} \right) \right) \right)^2
\end{equation}
the interference terms with a CLO are
\begin{align} \notag
&\mathrm{Int}_{jklm}^{\mathrm{CLO}} = \!\fracSumtwo{a_1=1}{A} \fracSumtwo{a_2=1}{A} \tilde{\lambda}_{jjk}^{(a_1)} \tilde{\lambda}_{jlm}^{(a_1)} \tilde{\lambda}_{jjk}^{(a_2)} \tilde{\lambda}_{jlm}^{(a_2)} \left( \tilde{\vect{x}}_{jk}^{\Htran}  \vect{D}_{\boldsymbol{\delta}(t)} \!\kron\! \vect{e}_{a_1}^{\Htran} \right) \\
& \! \times   \widetilde{\boldsymbol{\Psi}}^{-1}_j \left( \bar{\vect{X}}_{lm} \kron \vect{e}_{a_1} \vect{e}_{a_2}^{\Htran} \right) \widetilde{\boldsymbol{\Psi}}^{-1}_j \left(  \vect{D}_{\boldsymbol{\delta}(t)}^{\Htran}  \tilde{\vect{x}}_{jk} \!\kron\! \vect{e}_{a_2} \right) 
\end{align}
and the interference terms with SLOs are
\begin{align}
&\mathrm{Int}_{jklm}^{\mathrm{SLOs}} \!= \!
\left(\! \tr \!\left(\! \left( \tilde{\vect{x}}_{jk}^{\Htran}  \vect{D}_{\boldsymbol{\delta}(t)} \!\kron\! \tilde{\vect{\Lambda}}_{jjk}^{(A)} \right) \widetilde{\boldsymbol{\Psi}}^{-1}_j \left(  \vect{D}_{\boldsymbol{\delta}(t)}^{\Htran} \tilde{\vect{x}}_{lm}   \!\kron\! \tilde{\vect{\Lambda}}_{jlm}^{(A)}\right) \right)   \right)^2.
\end{align}
In these expresssions $\widetilde{\boldsymbol{\Psi}}_j \triangleq \sum_{\ell=1}^{L} \sum_{m=1}^{K} \vect{X}_{\ell m} \kron \tilde{\vect{\Lambda}}_{j \ell m}^{(A)}  + \xi \vect{I}_{AB}$,
$\vect{e}_{a}$ is the $a$th column of $\vect{I}_A$, and the big $\mathcal{O}$ notation $\mathcal{O}(\frac{1}{N})$ denotes terms that go to zero as $\frac{1}{N}$ or faster when $N \rightarrow \infty$.
\end{corollary}
\begin{IEEEproof}
The proof is given in Appendix E.
\end{IEEEproof}

This corollary shows that the distortion noise and receiver noise vanish as $N \rightarrow \infty$. The phase-drifts remain, but have no dramatic impact since these affect the numerator and denominator of the asymptotic SINR in \eqref{eq:asymptotic-SINR} in similar ways. The simulations in Section \ref{sec:numerical-results} show that the phase-drift degradations are not exacerbated in massive MIMO systems with SLOs, while the performance with a  CLO improves with $N$ but at a slower pace due to the phase-drifts.

The asymptotic SINRs are finite because both the signal power and parts of the inter-cell and intra-cell interference grow quadratically with $N$. This interference scaling behavior is due to so-called \emph{pilot contamination} (PC) \cite{Marzetta2010a,Jose2011b}, which represents the fact that a BS cannot fully separate signals from UEs that interfered with each other during pilot transmission.\footnote{Pilot contamination can be mitigated through semi-blind channel estimation as proposed in \cite{Mueller2014a}, but the UE rates will still be limited by hardware imperfections \cite{Bjornson2014a}.} Intra-cell PC is, conventionally, avoided by making the pilot sequences orthogonal in space; for example, by using the DFT pilot matrix $\widetilde{\vect{X}}_j^{\mathrm{spatial}}$ in Example \ref{example:pilot-sequences}. Unfortunately, the phase-drifts break any spatial pilot orthogonality. Hence, it is reasonable to remove intra-cell PC by assigning temporally orthogonal sequences, such as $\widetilde{\vect{X}}_j^{\mathrm{temporal}}$ in Example \ref{example:pilot-sequences}. Note that with temporal orthogonality the total pilot energy per UE, $\| \tilde{\vect{x}}_{jk} \|^2$, is reduced by $\frac{1}{K}$ since the energy per pilot symbol is constrained. Consequently, the simulations in Section \ref{sec:numerical-results} reveal that temporally orthogonal pilot sequences are only beneficial for extremely large arrays. Inter-cell PC cannot generally be removed, because there are only $B\leq T$ orthogonal sequences in the whole network, but it can be mitigated by allocating the same pilot to UEs that are well separated (e.g., in terms of second-order channel statistics such as different path-losses and spatial correlation \cite{Yin2013a}).

Apparently, the detrimental impact of hardware imperfections vanishes almost completely as $N$ grows large. This result holds for any fixed values of the parameters $\delta$, $\kappa$, and $\xi$. In fact, the hardware imperfections may even vanish when the hardware quality is gradually decreased with $N$. The next corollary formulates analytically such an important hardware scaling law.

\begin{corollary} \label{cor:scaling-law}
Suppose the hardware imperfection parameters are replaced as $\kappa^2 \mapsto \kappa_{0}^2 N^{z_1}$, $\xi \mapsto \xi_{0} N^{z_2}$, and $\delta \mapsto \delta_{0} (1+ \log_e(N^{z_3}) )$, for some given scaling exponents $z_1,z_2,z_3 \geq 0$ and some initial values $\kappa_0,\xi_0,\delta_0 \geq 0$. Moreover, let all pilot symbols be non-zero: $x_{jk}(\tau_b)>0$ for all $j$, $k$, and $b$. Then, all the SINRs, $\mathrm{SINR}_{jk}(t)$, under MRC receive filtering converge to non-zero limits as $N \rightarrow \infty$ if
\begin{equation} \label{eq:scaling-law}
\begin{cases} \max(z_1,z_2) \leq \frac{1}{2} \,\,\, \textrm{and} \,\,\, z_3=0 & \textrm{for a CLO} \\
\max(z_1,z_2) + z_3  \underset{\tau \in \{\tau_1,\ldots,\tau_B\}}{\min} \frac{\delta_{0} |t-\tau | }{2}   \leq \frac{1}{2} & \textrm{for SLOs}.
\end{cases}
\end{equation}
\end{corollary}
\begin{IEEEproof}
The proof is given in Appendix F.
\end{IEEEproof}

This corollary proves that we can tolerate stronger hardware imperfections as the number of antennas increases. This is a very important result for practical deployments, because we can relax the design constraints on the hardware quality as $N$ increases. In particular, we can achieve better energy efficiency in the circuits and/or lower hardware costs by accepting larger distortions than conventionally. This property has been conjectured in overview articles, such as \cite{Larsson2014a}, and was proved in \cite{Bjornson2014a} using a simplified system model with only additive distortion noise. Corollary \ref{cor:scaling-law} shows explicitly that the conjecture is also true for multiplicative phase-drifts, receiver noise, and inter-carrier interference. Going a step further, Section \ref{sec:circuit-examples} exemplifies how the scaling law may impact the circuit design in practical deployments.

Since Corollary \ref{cor:scaling-law} is derived for MRC filtering, \eqref{eq:scaling-law} provides a \emph{sufficient} scaling condition also for any receive filter that performs better than MRC. The scaling law for SLOs consists of two terms: $\max(z_1,z_2)$ and $z_3  \underset{\tau \in \{\tau_1,\ldots,\tau_B\}}{\min} \frac{\delta_{0} |t-\tau | }{2}$. The first term $\max(z_1,z_2)$ shows that the additive distortion noise and receiver noise can be increased simultaneously and independently (as fast as $\sqrt{N}$), while the sum of the two terms manifests a tradeoff between allowing hardware imperfections that cause additive and multiplicative distortions. The scaling law for a CLO  allows only for increasing the additive distortion noise and receiver noise, while the phase-drift variance should not be increased because only the signal gain (and not the interference) is reduced by phase-drifts in this case; see Example \ref{example:CLOvsSLOs}. Clearly, the system is particularly vulnerable to phase-drifts due to their accumulation and since they affect the signal itself; even in the case of SLOs, the second term of \eqref{eq:scaling-law} increases with $T$ and the variance $\delta$ can scale only logarithmically with $N$. Note that we can accept larger phase-drift variances if the coherence block $T$ is small and the pilot symbols are distributed over the coherence block, which is in line with the results in \cite{Pitarokoilis2014a}.

\newpage

\section{Utilizing the Scaling Law: \\ Circuit-Aware Design}
\label{sec:circuit-examples}

The generic system model with hardware imperfections in \eqref{eq:generalized-model} describes a flat-fading multi-cell channel. This channel can describe either single-carrier transmission over the full available flat-fading bandwidth as in \cite{Mehrpouyan2012a} or one of the subcarriers in a system based on multi-carrier modulation; for example, OFDM or FBMC as in \cite{Schenk2008a,Farhang2014a}. To some extent, it can also describe single-carrier transmission over frequency-selective channels as in \cite{Pitarokoilis2014a}. The mapping between the imperfections in a certain circuit in the receiving array to the three categories of distortions (defined in Section \ref{sec:system-model}) depends on the modulation scheme. For example, the multiplicative distortions caused by phase-noise leads also to inter-carrier interference in OFDM which is an additive noise-like distortion.

In this section, we exemplify what the scaling law in Corollary \ref{cor:scaling-law} means for the circuits depicted in Fig.~\ref{figure_typicalreceiver}. In particular, we show that the scaling law can be utilized for circuit-aware system design, where the cost and power dissipation per circuit will be gradually decreased to achieve a sub-linear cost/power scaling with the number of antennas. For clarity of presentation, we concentrate on single-carrier transmission over flat-fading channels, but mention briefly if the interpretation might change for multi-carrier modulation.

\subsection{Analog-to-Digital Converter (ADC)}

The ADC quantizes the received signal to a $b$ bit resolution. Suppose the received signal power is $P_{\textrm{signal}}$ and that automatic gain control is used to achieve maximum quantization accuracy irrespective of the received signal power. In terms of the originally received signal power $P_{\textrm{signal}}$, the quantization in single-carrier transmission can be modeled as reducing the signal power to $(1-2^{-2b}) P_{\textrm{signal}}$ and adding uncorrelated quantization noise with power $2^{-2b} P_{\textrm{signal}}$ \cite[Eq.~(17)]{Mezghani2010a}. This model is particularly accurate for high ADC resolutions.
 We can include the quantization noise in the channel model \eqref{eq:generalized-model} by normalizing the useful signal. The quantization noise is included in the additive distortion noise $\boldsymbol{\upsilon}_{j}(t)$ and contributes to $\kappa^2$ with $\frac{2^{-2b}}{1-2^{-2b}}$, while the receiver noise variance $\xi$ is scaled by a factor $\frac{1}{1-2^{-2b}}$ due to the normalization.
The scaling law in Corollary \ref{cor:scaling-law}  allows us to increase the variance $\kappa^2$ as $N^{z_1}$ for $z_1 \leq \frac{1}{2}$. This corresponds to reducing the ADC resolution by around $\frac{z_1}{2} \log_2(N)$ bits, which reduces cost and complexity. For example, we can reduce the ADC resolution per antenna by 2 bits if we deploy 256 antennas instead of one. For very large arrays, it is even sufficient to use 1-bit ADCs (cf.~\cite{Risi2014a}).

The power dissipation of an ADC, $P_{\mathrm{ADC}}$, is proportional to $2^{2b}$ \cite[Eq.~(14)]{Mezghani2010a} and can, thus, be decreased approximately as $1/N^{z_1}$. If each antenna has a separate ADC, the total power $N P_{\mathrm{ADC}}$ increases with $N$ but proportionally to $N^{1-z_1}$, for  $z_1 \leq \frac{1}{2}$, instead of $N$, due to the gradually lower ADC resolution. The scaling can thus be made as small as $\sqrt{N}$.

\subsection{Low Noise Amplifier (LNA)}

The LNA is an analog circuit that amplifies the received signal. It is shown in \cite{Song2008b} that the behavior of an LNA is characterized by the figure-of-merit (FoM) expression
\begin{equation} \label{eq:FOM-LNA}
\mathrm{FoM}_{\mathrm{LNA}} = \frac{G}{(F-1) P_{\mathrm{LNA}}}
\end{equation}
where $F \geq 1$ is the noise amplification factor, $G$ is the amplifier gain, and $P_{\mathrm{LNA}}$ is the power dissipation in the LNA.
Using this notation, the LNA contributes to the receiver noise variance $\xi$ with $F \sigma^2$. For optimized LNAs, $\mathrm{FoM}_{\mathrm{LNA}}$ is a constant determined by the circuit architecture \cite{Song2008b}; thus, $\mathrm{FoM}_{\mathrm{LNA}}$ basically scales with the hardware cost. The scaling law in Corollary \ref{cor:scaling-law} allows us to increase $\xi$ as  $N^{z_2}$ for $z_2 \leq \frac{1}{2}$. The noise figure, defined as $10 \log_{10} (F)$, can thus be increased by $z_2 10 \log_{10} (N)$ dB. For example, at $z_2 = \frac{1}{2}$ we can allow an increase by 10 dB if we deploy 100 antennas instead of one.

For a given circuit architecture, the invariance of the $\mathrm{FoM}_{\mathrm{LNA}}$ in \eqref{eq:FOM-LNA} implies that we can decrease the power dissipation (roughly) proportional to $1/N^{z_2}$. Hence, we can make the total power dissipation of the $N$ LNAs, $N P_{\mathrm{LNA}}$, increase as $N^{1-z_2}$ instead of $N$ by tolerating higher noise amplification. The scaling can thus be made as small as $\sqrt{N}$.

\subsection{Local Oscillator (LO)}

Phase noise in the LOs is the main source of multiplicative phase-drifts and changes the phases gradually at each channel use. The average amount of phase-drifts that occurs under a coherence block is $\delta T$ and depends on the phase-drift variance $\delta$ and the block length $T$. If the LOs are free-running, the phase noise is commonly modeled by the Wiener process (random walk) defined in Section \ref{sec:system-model} \cite{Petrovic2007a,Mehrpouyan2012a,Durisi2013b,Pitarokoilis2014a,Petrovic2004a} and the phase noise variance is given by 
\begin{equation} \label{eq:oscillator-variance}
\delta = 4\pi^2 f_c^2 T_s \zeta
\end{equation} 
where $f_c$ is the carrier frequency, $T_s$ is the symbol time, and $\zeta$ is a constant that characterizes the quality of the LO  \cite{Petrovic2007a}. If $\delta$ and/or $T$ are small, such that $\delta T \approx 0$, the channel variations dominate over the phase noise. However, phase noise can play an important role when modeling channels with large coherence time (e.g., fixed indoor users, line-of-sight, etc.) and as the carrier frequency increases (since $\delta = \mathcal{O}(f_c^2)$ while the Doppler spread reduces $T$ as $\mathcal{O}(f_c^{-1}$) \cite{Mehrpouyan2012a}. Relevant examples are mobile broadband access to homes and WiFi at millimeter frequencies.

The power dissipation $P_{\mathrm{LO}}$ of the LO is coupled to $\zeta$, such that $P_{\mathrm{LO}} \zeta \approx \mathrm{FoM}_{\mathrm{LO}}$ where the FoM value $\mathrm{FoM}_{\mathrm{LO}}$ depends on the circuit architecture \cite{Petrovic2007a,Park2008a} and naturally on the hardware cost. For a given architecture, we can allow larger $\delta$ and, thereby, decrease the power $P_{\mathrm{LO}}$. The scaling law in Corollary \ref{cor:scaling-law} allows us to increase $\delta$ as $(1+ \log_e(N^{z_3}) )$ when using SLOs. The power dissipation per LO can then be reduced as $\frac{1}{1+z_3 \log_e(N)}$. This reduction is only logarithmic in $N$, which stands in contrast to the $1/\sqrt{N}$ scalings for ADCs and LNAs (achieved by $z_1=z_2 =\frac{1}{2}$). Since linear increase is much faster than logarithmic decay, the total power $N P_{\mathrm{LO}}$ with SLOs increases almost linearly with $N$; thus, the benefit is mostly cost and design related. In contrast, the phase noise variance cannot be scaled when having a CLO, because massive MIMO only relaxes the design of circuits that are placed independently at each antenna branch.

Imperfections in the LOs also cause inter-carrier interference in OFDM systems, since the subcarrier orthogonality  is broken \cite{Schenk2008a}. When inter-carrier interference is created at the receiver side it depends on the channels of other subcarriers. It is thus uncorrelated with the useful channel in \eqref{eq:generalized-model} and can be included in the receiver noise term. Irrespective of the type of LOs, the severity of inter-carrier interference is suppressed by $z_2 10 \log_{10} (N)$ dB according to Corollary \ref{cor:scaling-law}. Hence, massive MIMO is less vulnerable to in-band distortions than conventional systems.

The phase-noise variance formula in \eqref{eq:oscillator-variance} gives other possibilities than decreasing the circuit power. In particular, one can increase the carrier frequency $f_c$ with $N$ by using Corollary \ref{cor:scaling-law}. This is an interesting observation since massive MIMO has been identified as a key enabler for millimeter-wave communications \cite{Baldemair2013a}, in which the phase noise is more severe since the variance in \eqref{eq:oscillator-variance} increases quadratically with the carrier frequency $f_c$. Fortunately, massive MIMO with SLOs has an inherent resilience to phase noise.

\subsection{Non-Linearities}

Although the physical propagation channel is linear, practical systems can exhibit non-linear behavior due to a variety of reasons; for examples, non-linearities in filters, converters, mixers, and amplifiers \cite{Schenk2008a} as well as passive intermodulation caused by various electro-thermal phenomena \cite{Wilkerson2010a}. Such non-linearities are often modeled by power series or Volterra series \cite{Wilkerson2010a}, but since we consider a system with Gaussian transmit signals the Bussgang theorem can be applied to simplify the characterization \cite{Holma2011a,Zhang2012a}. For a Gaussian variable $X$ and any non-linear function $g(\cdot)$, the Bussgang theorem implies that $g(X) = c X + V$, where $c$ is a scaling factor and $V$ is a distortion uncorrelated with $X$; see \cite[Eq.~(15)]{Zhang2012a}. If we let $g(X)$ describe a nonlinear component and let $X$ be the useful signal, the impact of non-linearities can be modeled by a scaling of the useful signal and an additional distortion term. Depending on the nature of each non-linearity, the corresponding distortion is either included in the distortion noise or the receiver noise.\footnote{The distortion from non-linearities are generally non-Gaussian, but this has no impact on our analysis because the achievable rates in Lemma \ref{lemma:achievable-rates} were obtained by making the worst-case assumption of all additive distortions being Gaussian distributed.} The scaling factor $c$ of the useful signal is removed by scaling $\kappa^2$ and $\xi$ by $\frac{1}{|c|^2}$.

\section{Numerical Illustrations}
\label{sec:numerical-results}

\begin{figure*}[t!]
\begin{center}
\includegraphics[width=1.6\columnwidth]{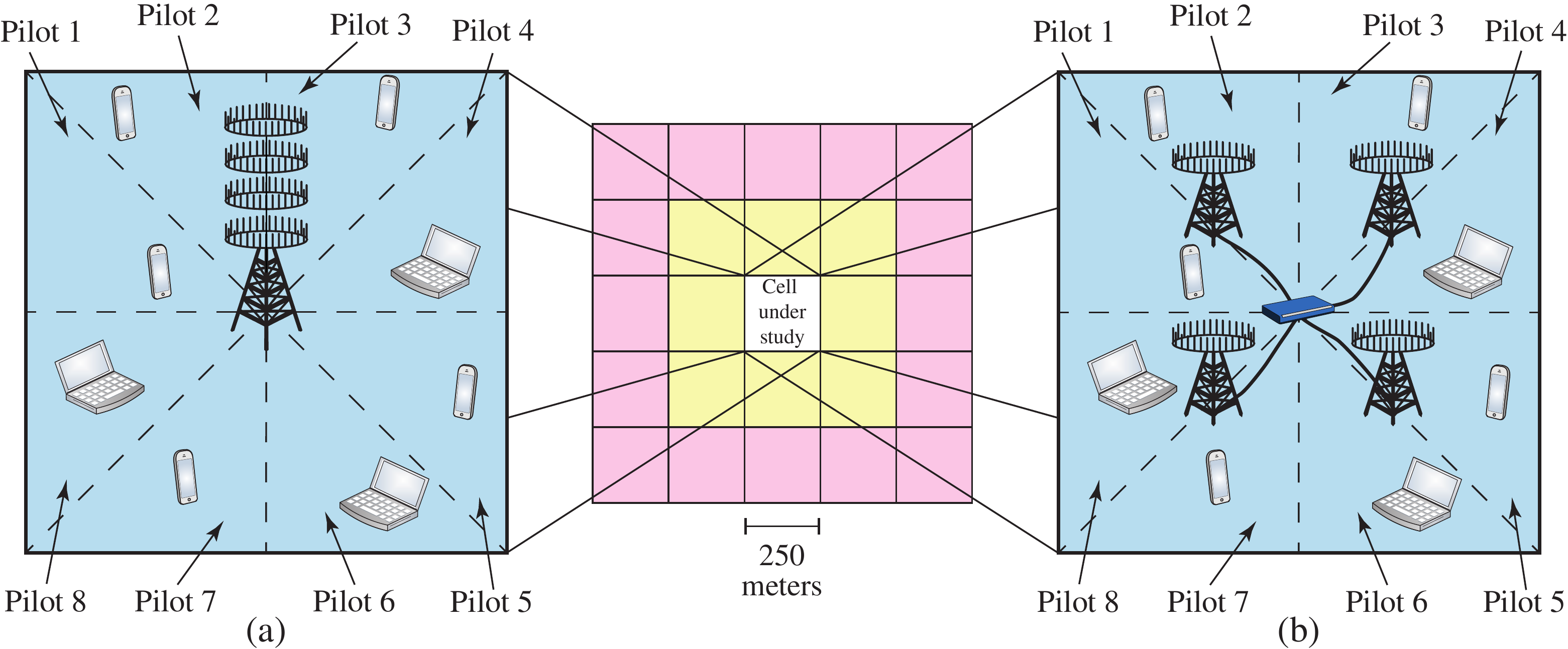}
\end{center} \vskip-3mm
\caption{The simulations consider the uplink of a cell surrounded by two tiers of interfering cells. Each cell contains $K=8$ UEs that are uniformly distributed in different parts of the cell. Two site deployments are considered: (a) $N$ co-located antennas in the middle of the cell; and (b) $N/4$ antennas at 4 distributed arrays.} \label{figure_simulation-scenario}
\end{figure*}

Our analytic results are corroborated in this section by studying the uplink in a cell surrounded by 24 interfering cells, as shown in Fig.~\ref{figure_simulation-scenario}. Each cell is a square of $250 \,\text{m} \times  250\,\text{m}$ and we compare two topologies: (a) co-located deployment of $N$ antennas in the middle of the cell; and (b) distributed deployment of 4 subarrays of $\frac{N}{4}$ antennas at distances of $62.5 \,\text{m}$ from the cell center. To mimic a simple user scheduling algorithm, each cell is divided into 8 virtual sectors and one UE is picked with a uniform distribution in each sector (with a minimum distance of $25\,\text{m}$ from any array location). We thus have $K=8$ and use $B=8$ as pilot length in this section.
Each sector is allocated an orthogonal pilot sequence, while the same pilot is reused in the same sector of all other cells.
The channel attenuations are modeled as \cite{LTE2010b}
\begin{equation}
\lambda_{jlk}^{(n)} = \frac{10^{s_{jlk}^{(n)}-1.53}}{(d_{jlk}^{(n)})^{3.76}}
\end{equation}
where $d_{jlk}^{(n)}$ is the distance in meters between receive antenna $n$ in cell $j$ and UE $k$ in cell $l$ and $s_{jlk}^{(n)} \sim \mathcal{N}(0,3.16)$ is shadow-fading  (it is the same for co-located antennas but independent between the 4 distributed arrays). We consider statistical power control with $p_{jk} = \frac{\rho}{\frac{1}{N} \sum_{n=1}^{N} \lambda_{jjk}^{(n)}}$ to achieve an average received signal power of $\rho$ over the receive antennas. The thermal noise variance is $\sigma^2 = -174$ dBm/Hz. We consider average SNRs, $\rho/\sigma^2$, of 5 and 15 dB, leading to reasonable transmit powers (below 200 mW over a 10 MHz bandwidth) for UEs at cell edges. The simulations were performed using Matlab and the code is available for download at \url{https://github.com/emilbjornson/hardware-scaling-laws}, which enables reproducibility as well as simple testing of other parameter values.

\subsection{Comparison of Deployment Scenarios}

We first compare the co-located and distributed deployments in Fig.~\ref{figure_simulation-scenario}.  We consider the MRC filter, set the coherence block to $T=500$ channel uses (e.g., 5 ms coherence time and 100 kHz coherence bandwidth), use the DFT-based pilot sequences of length $B=8$, and send these in the beginning of the coherence block. The results are averaged over different UE locations.

\begin{figure}[t!]
\begin{center}
\includegraphics[width=\columnwidth]{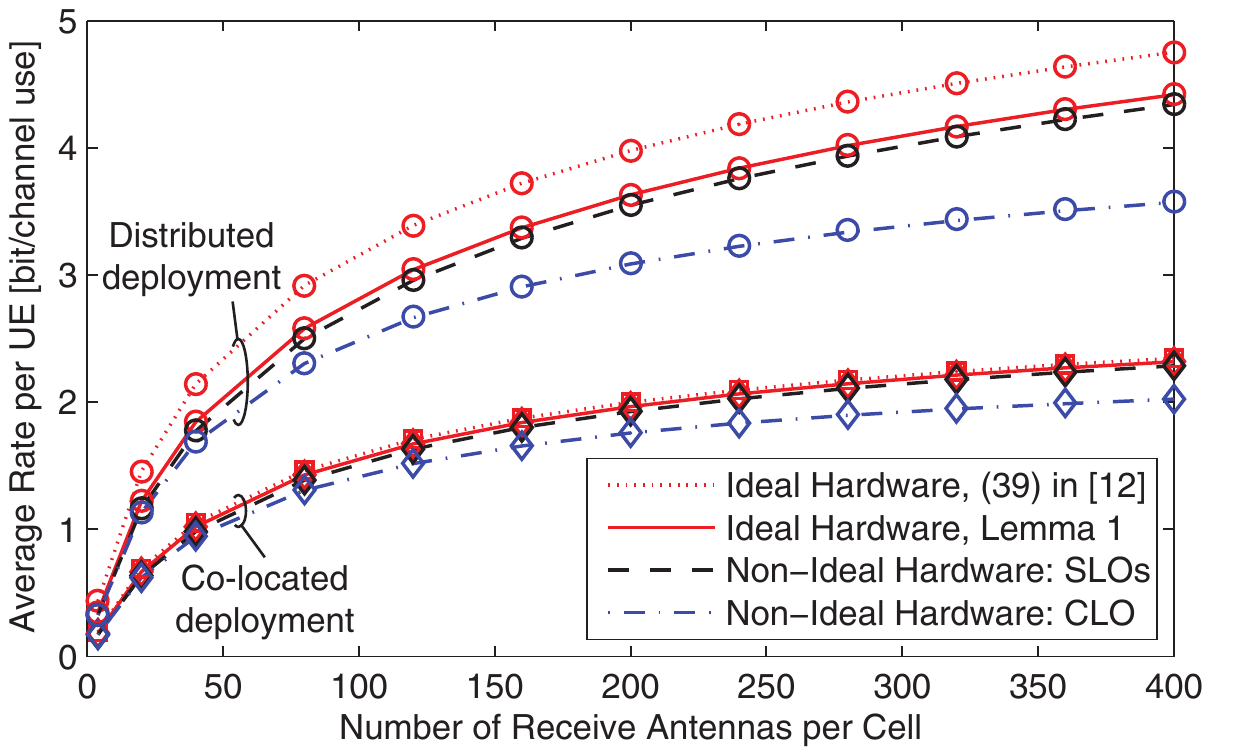}
\end{center} \vskip-3mm
\caption{Achievable rates with MRC filter and either ideal hardware or imperfections given by $(\kappa,\xi,\delta) = (0.0156, \, 1.58 \sigma^2, \, 1.58 \cdot 10^{-4} )$. Co-located and distributed antenna deployments are compared, as well as, a CLO and SLOs.} \label{figure_scenarios_mc}
\end{figure}

The average achievable rates per UE are shown in Fig.~\ref{figure_scenarios_mc} for $\rho/\sigma^2 = 5$ dB, using either ideal hardware or imperfect hardware with $\kappa = 0.0156$, $\xi =  1.58 \sigma^2$, and $\delta = 1.58 \cdot 10^{-4}$. These parameter values were not chosen arbitrarily, but based on the circuit examples in Section \ref{sec:circuit-examples}. More specifically, we obtained $\kappa = 2^{-b}/\sqrt{1-2^{-2b}} = 0.0156$ by using $b = 6 \, \mathrm{bit}$ ADCs and $\xi = \frac{F \sigma^2}{1-2^{-2b}} = 1.58 \sigma^2$ for a noise amplification factor of $F = 2$ dB. The phase noise variance $\delta = 1.58 \cdot 10^{-4}$ was obtained from \eqref{eq:oscillator-variance} by setting $f_c = 2 \, \mathrm{GHz}$, $T_s = 10^{-7} \, \mathrm{s}$, and $\zeta = 10^{-17}$. Note that the curves in Fig.~\ref{figure_scenarios_mc} are based on the analytic results in Theorem \ref{theorem:MRC-expectations}, while the marker symbols correspond to Monte Carlo simulations of the expectations in \eqref{eq:achievable-SINR}. The perfect match validates the analytic results.

Looking at Fig.~\ref{figure_scenarios_mc}, we see that the tractable ergodic rate from Lemma \ref{lemma:achievable-rates} approaches well the slightly higher achievable rate from \cite[Eq.~(39)]{Ngo2013a}. Moreover, we see that the hardware imperfections cause small rate losses when the number of antennas, $N$, is small. However, the large-$N$ behavior depends strongly on the oscillators: the rate loss is small for SLOs at any $N$, while it can be very large if a CLO is used when $N$ is large (e.g., 25\% rate loss at $N=400$). This important property was explained in Example \ref{example:CLOvsSLOs} and the simple explanation is that the effect of phase noise averages out with SLOs, but at the cost of adding more hardware.

Fig.~\ref{figure_scenarios_mc} also shows that the distributed massive MIMO deployment achieves roughly twice the rates of co-located massive MIMO. This is because distributed arrays can exploit both the proximity gains (normally achieved by small cells) and the array gains and spatial resolution of coherent processing over many antennas. 

\begin{figure}[t!]
\begin{center}
\includegraphics[width=\columnwidth]{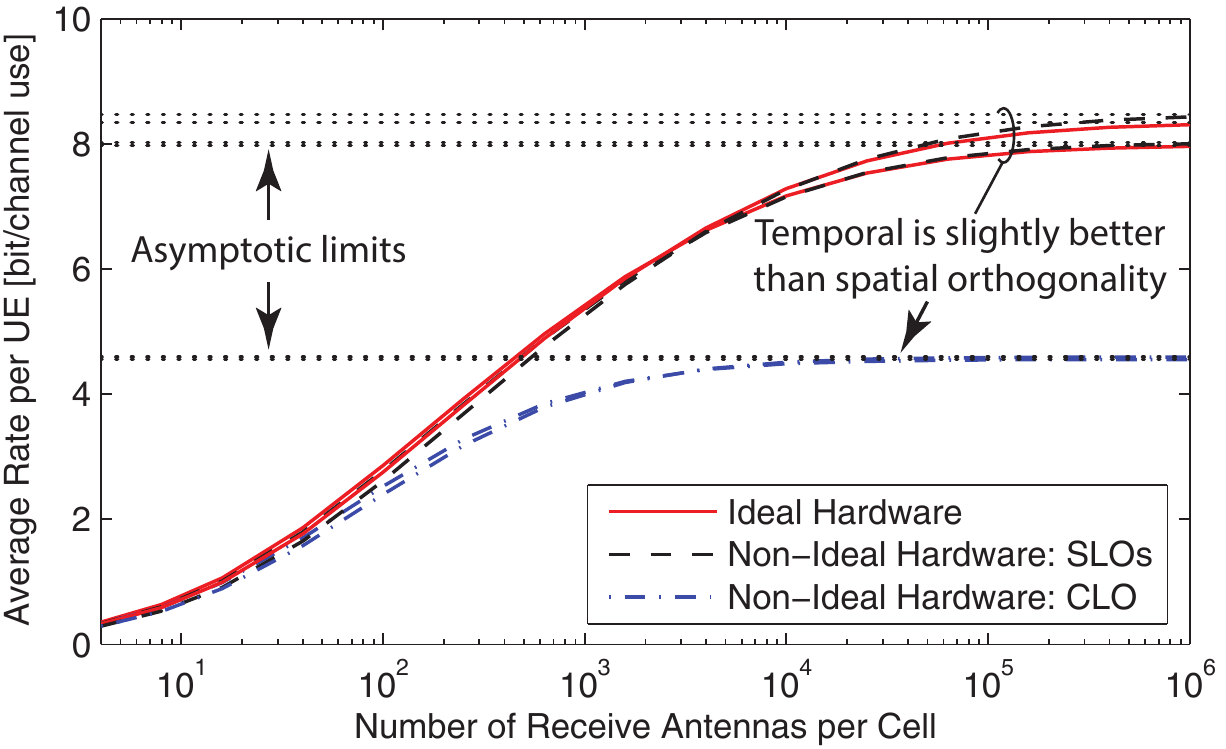}
\end{center} \vskip-3mm
\caption{Average UE rate with MRC filter for different numbers of antennas, different hardware imperfections, and spatially or temporally orthogonal pilots. Note the logarithmic horizontal scale which is used to demonstrate the asymptotic behavior.} \label{figure_asymptotics}
\end{figure}

\subsection{Validation of Asymptotic Behavior}

Next, we illustrate the asymptotic behavior of the UE rates (with MRC filter) as $N \rightarrow \infty$. For the sake of space, we only consider the distributed deployment in Fig.~\ref{figure_simulation-scenario}, while a similar figure for the co-located deployment is available in \cite{Bjornson2014c}. Fig.~\ref{figure_asymptotics} shows the UE rates as a function of the number of antennas, for ideal hardware and the same hardware imperfections as in the previous figure. The simulation validates the convergence to the limits derived in Corollary \ref{corollary:asymptotic-SINR}, but also shows that the convergence is very slow---we used logarithmic scale on the horizontal axis because $N=10^6$ antennas are required for convergence for ideal hardware and for hardware imperfections with SLOs, while $N=10^4$ antennas are required for hardware imperfections with a CLO. The performance loss for hardware imperfections with SLOs is almost negligible, while the loss when having a CLO grows with $N$ and approaches $50 \, \%$.

Two types of pilot sequences are also compared in Fig.~\ref{figure_asymptotics}: the temporally orthogonal pilots in \eqref{eq:temporal-orthogonal-matrix} and the spatially orthogonal DFT-based pilots in \eqref{eq:dft-pilot-matrix}. As discussed in relation to Corollary \ref{corollary:asymptotic-SINR}, temporal orthogonality provides slightly higher rates in the asymptotic regime (since the phase noise cannot break the temporal pilot orthogonality). However, this gain is barely visible in Fig.~\ref{figure_asymptotics} and only kicks in at impractically large $N$. Since temporally orthogonal pilots use $K$ times less pilot energy, they are the best choice in this simulation. However, if the average SNR is decreased then spatially orthogonal pilots can be used to improve the estimation accuracy.

\subsection{Impact of Coherence Block Length}

\begin{figure}[t!]
\begin{center}
\includegraphics[width=\columnwidth]{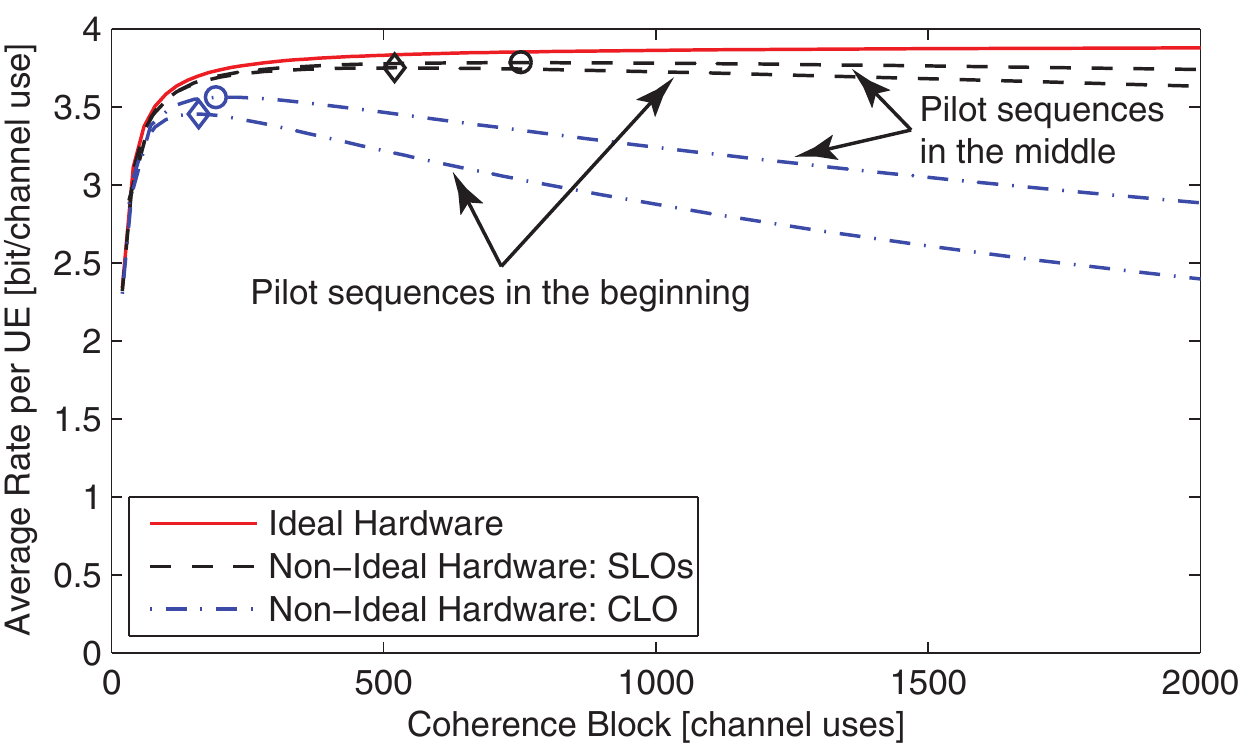}
\end{center} \vskip-3mm
\caption{Average UE rate with MRC filter as a function of the coherence block length, for different pilot sequence distributions. The maximum is marked at each curve and is the preferable operating point for the transmission protocol.} \label{figure_coherenceblock}
\end{figure}

Next, we illustrate how the length of the coherence block, $T$, affects the UE rates with the MRC filter. We consider a practical number of antennas, $N = 240$, while having $\rho/\sigma^2 = 5$ dB and imperfections with $\{ \kappa, \, \xi, \, \delta \} \!=\! \{ 0.0156, \, 1.58 \sigma^2, \, 1.58 \cdot 10^{-4}\}$, as before. The UE rates are shown in Fig.~\ref{figure_coherenceblock} as a function of $T$. We compare two ways of distributing the pilot sequences over the coherence block: in the beginning or in the middle (see (a) and (b) in Fig.~\ref{figure_protocol}).

With ideal hardware, the pilot distribution has no impact. We observe in Fig.~\ref{figure_coherenceblock} that the average UE rates are slightly increasing with $T$. This is because the pre-log penalty of using only $T-B$ out of $T$ channel uses for data transmission is smaller when $T$ is large (and $B$ is fixed). In the case with hardware imperfections, Fig.~\ref{figure_coherenceblock} shows that the rates increase with $T$ for small $T$ (for the same reason as above) and then decrease with $T$ since phase-drifts accumulate over time.

Interestingly, slightly higher rates are achieved and larger coherence blocks can be handled if the pilot sequences are sent in the middle of the coherence block (instead of the beginning) since the phase drifts only accumulate half as much. From an implementation perspective it is, however, better to put pilot sequences in the beginning, since then there is no need to buffer the incoming signals while waiting for the pilots that enable computation of receive filters.

Fig.~\ref{figure_coherenceblock} shows, once again, that systems with SLOs have higher robustness to phase-drifts than systems with a CLO. To make a fair comparison, we need to consider that the coherence block is a modeling concept---we can always \emph{choose} a transmission protocol with a smaller $T$ than prescribed by the coherence block length, at the cost of increasing the pilot overhead $B/T$. Hence, it is the maximum at each curve, indicated by markers, which is the operating point to compare. The difference between SLOs and a CLO is much smaller when comparing the maxima, but these are achieved at very different $T$-values; the transmission protocol should send pilots much more often when having a CLO. The true optimum is achieved by maximizing over $T$ and $B$, and probably by spreading the pilots to reduce the accumulation of phase drifts.

\subsection{Hardware Scaling Laws with Different Receive Filters}

\begin{figure}[t!]
\begin{center}
\includegraphics[width=\columnwidth]{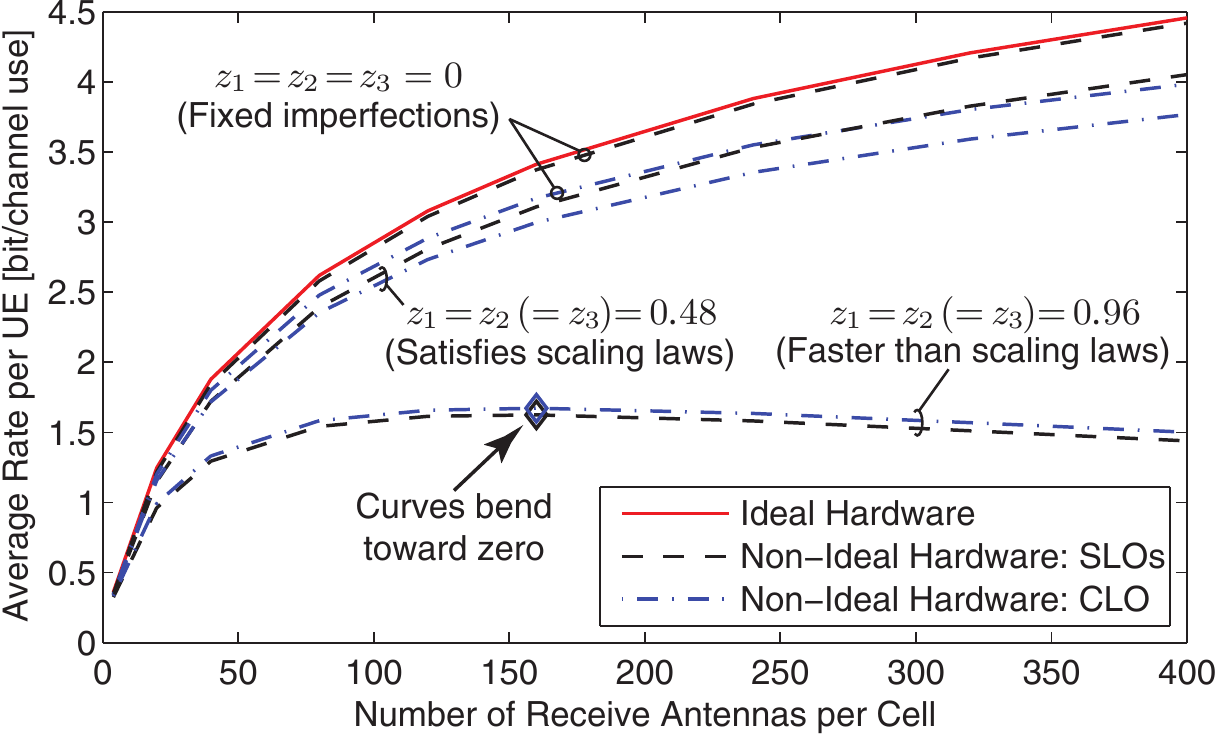}
\end{center} \vskip-3mm
\caption{Average UE rate with MRC filter for different numbers of antennas, $N$, and with hardware imperfections that are either fixed or increase with $N$ as in Corollary \ref{cor:scaling-law}.} \label{figure_scalinglaw} 
\end{figure}

\begin{figure}[t!]
\begin{center}
\includegraphics[width=\columnwidth]{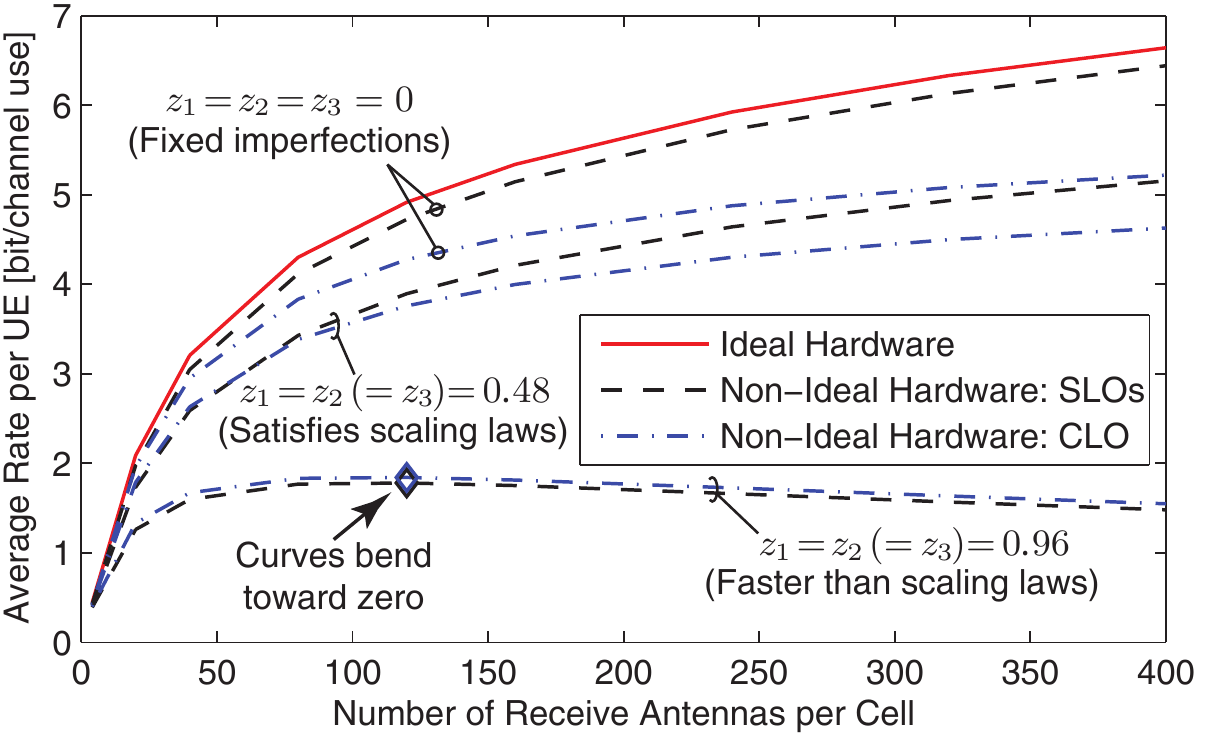}
\end{center} \vskip-3mm
\caption{Average UE rate with MMSE filter (in \eqref{eq:MMSE-receiver}) for different numbers of antennas, $N$, and with hardware imperfections that are either fixed or increase with $N$ as in Corollary \ref{cor:scaling-law}.} \label{figure_scalinglaw_MMSE}
\end{figure}

Next, we illustrate the scaling laws for hardware imperfections established in Corollary \ref{cor:scaling-law} and set $\rho/\sigma^2 \!=\! 15$ dB to emphasize this effect. We focus on the distributed scenario for $T=500$, since the co-located scenario behaves similarly and can be found in \cite{Bjornson2014d}. Using the notation from the scaling law, we set the baseline hardware imperfections to $( \kappa_0, \, \xi_0, \, \delta_0 ) \!=\! ( 0.05, \, 3 \sigma^2, \, 7 \cdot 10^{-5} )$ and increase these with $N$ using different values on the exponents $z_1$, $z_2$, and $z_3$.

The UE rates with MRC filters are given in Fig.~\ref{figure_scalinglaw} for ideal hardware, fixed hardware imperfections, and imperfections that either increase according to the scaling law or faster than the law (observe that we always have $z_3 = 0$ for a CLO). As expected, the $z$-combinations that satisfy the scaling laws give small performance losses, while the bottom curves go asymptotically to zero since the law is not fulfilled (the points where the curves bend downwards are marked).

We have considered the MRC filter since its low computational complexity is attractive for massive MIMO topologies. MRC provides a performance baseline for other receive filters which typically have higher complexity.  In Fig.~\ref{figure_scalinglaw_MMSE} we consider the (approximate) MMSE receive filter
\begin{align} \notag
&\vect{v}_{jk}^{\textrm{MMSE}}(t) \\
&\triangleq \left( \sum_{l=1}^{L} \sum_{m=1}^{K} p_{lm} \big(\vect{G}_{jlm}(t) \!+\! \kappa^2 \vect{D}_{\vect{G}_{jlm}(t)} \big)  \!+\! \xi \vect{I}_M  \! \right)^{\!\!-1} \!\! \hat{\vect{h}}_{jjk}(t)  \label{eq:MMSE-receiver}
\end{align} 
where $\vect{G}_{jlm}(t) \triangleq \hat{\vect{h}}_{jlm}(t)  \hat{\vect{h}}_{jlm}^{\Htran}(t) + \vect{C}_{jlm}(t)$ and $\vect{D}_{\vect{G}_{jlm}(t)}$ is a diagonal matrix where the diagonal elements are the same as in $ \vect{G}_{jlm}(t)$. By comparing Figs.~\ref{figure_scalinglaw} and \ref{figure_scalinglaw_MMSE} (notice the different scales), we observe that the MMSE filter provides higher rates than the MRC filter. Interestingly, the losses due to hardware imperfections behave similarly but are larger for MMSE. This is because the MMSE filter exploits spatial interference suppression which is  sensitive to imperfections.

\section{Conclusion}
\label{sec:conclusion}

Massive MIMO technology can theoretically improve the spectral and energy efficiencies by orders of magnitude, but to make it a commercially viable solution it is important that the $N$ antenna branches can be manufactured using low-cost and low-power components. As exemplified in Section \ref{sec:circuit-examples}, such components are prone to hardware imperfections that distort the communication and limit the achievable performance.

In this paper, we have analyzed the impact of such hardware imperfections at the BSs by studying an uplink communication model with multiplicative phase-drifts, additive distortion noise, noise amplifications, and inter-carrier interference. The system model can be applied to both co-located and distributed antenna arrays. We derived a new LMMSE channel estimator/predictor and the corresponding achievable UE rates with MRC. Based on these closed-form results, we prove that only the phase-drifts limit the achievable rates as $N \rightarrow \infty$. This showcases that massive MIMO systems are robust to hardware imperfections, which is a property that has been conjectured in prior works (but only proved for simple models with one type of imperfection). This phenomenon can be attributed to the fact that distortions are uncorrelated with the useful signals and, thus, add non-coherently during the receive processing.

Particularly, we established a scaling law showing that the variance of the distortion noise and receiver noise can increase simultaneously as $\sqrt{N}$. If the phase-drifts are independent between the antennas, we can also tolerate an increase of the phase-drift variance with $N$, but only logarithmically. If the phase-drifts are the same over the antennas (e.g., if a CLO is used), then the phase-drift variance cannot increase. The numerical results show that there are substantial performance benefits of using separate oscillators at each antenna branch instead of a common oscillator. The difference in performance might be smaller if the LMMSE estimator is replaced by a Kalman filter that exploits the exact distribution of the phase-drifts \cite{Bjornson2014a,Krishnan2015a}. Interestingly, the benefit of having SLOs remains also under idealized uplink conditions (e.g., perfect CSI, no interference, and high SNR \cite{Khanzadi2015a}).
In any case, the transmission protocol must be adapted to how fast the phase-drifts deteriorate performance. The scaling law was derived for MRC but provides a sufficient condition for other judicious receive filters, like the MMSE filter.
We also exemplified what the scaling law means for different circuits in the receiver (e.g., ADCs, LNAs, and LOs). This quantifies how fast the requirements on the number of quantization bits and the noise amplification can be relaxed with $N$. It also shows that a circuit-aware design can make the total circuit power consumption of the $N$ ADCs and LNAs increase as $\sqrt{N}$, instead of $N$ which would conventionally be the case.

A natural extension to this paper would be to consider also the downlink with hardware imperfections at the BSs.
If maximum ratio transmission (MRT) is used for precoding, then more-or-less the same expectations as in the uplink SINRs in \eqref{eq:achievable-SINR} will show up in the downlink SINRs but at different places \cite{Bjornson2014lecturenote}. Hence, we believe that similar closed-form rate expressions and scaling laws for the levels of hardware imperfections can be derived. The analytic details and interpretations are, however, outside the scope of this paper.

\appendices

\section*{Appendix A: A Useful Lemma}

\begin{lemma} \label{lemma:second-order-moment-Gaussian}
Let $\vect{u} \sim \mathcal{CN}(\vect{0},\vect{\Lambda})$ and consider some deterministic matrix $\vect{M}$. It holds that
\begin{equation}
\mathbb{E} \left\{ | \vect{u}^{\Htran} \vect{M} \vect{u} |^2 \right\} =  |\tr( \vect{\Lambda} \vect{M})|^2 + \tr( \vect{\Lambda} \vect{M} \vect{\Lambda} \vect{M}^{\Htran}).
\end{equation}
\end{lemma}
\begin{IEEEproof}
This lemma follows from straightforward computation, by exploiting that $\vect{u}^{\Htran} \vect{M} \vect{u} = \tilde{\vect{u}}^{\Htran} \vect{\Lambda}^{1/2} \vect{M} \vect{\Lambda}^{1/2}\tilde{\vect{u}}
= \sum_{i,j} \tilde{u}_i^*  [ \vect{\Lambda}^{1/2} \vect{M} \vect{\Lambda}^{1/2} ]_{i,j} \tilde{u}_j$ where $\tilde{\vect{u}} \sim \mathcal{CN}(\vect{0},\vect{I})$.
\end{IEEEproof}

\section*{Appendix B: Proof of Theorem \ref{theorem:LMMSE-estimation}}
\label{proof:theorem:LMMSE-estimation}

We exploit the fact that $\hat{\vect{h}}_{jlk}(t) = \mathbb{E}\{ \vect{h}_{jlk}(t) \boldsymbol{\psi}_j^{\Htran} \}  \left( \mathbb{E}\{ \boldsymbol{\psi}_j \boldsymbol{\psi}_j^{\Htran} \}  \right)^{-1}  \boldsymbol{\psi}_j$ is the general expression of an LMMSE estimator \cite[Ch.~12]{Kay1993a}. Since the additive distortion and receiver noises are uncorrelated with $\vect{h}_{jlk}(t)$ and the UEs' channels are independent, we have that
\begin{align} \notag
&\mathbb{E}\{ \vect{h}_{jlk}(t) \boldsymbol{\psi}_j^{\Htran} \} \\ \notag & \!= \mathbb{E}\{ \vect{D}_{\boldsymbol{\phi}_j(t)} \vect{h}_{jlk} \vect{h}_{jlk}^{\Htran} [  \vect{D}_{\boldsymbol{\phi}_j(\tau_1)}^{\Htran} x_{lk}^*(\tau_1) \, \ldots \, \vect{D}_{\boldsymbol{\phi}_j(\tau_B)}^{\Htran} x_{lk}^*(\tau_B) ] \} \\ \notag
& \!= \vect{\Lambda}_{jlk}  \mathbb{E}\{  [ \vect{D}_{\boldsymbol{\phi}_j(t)} \vect{D}_{\boldsymbol{\phi}_j(\tau_1)}^{\Htran} x_{lk}^*(\tau_1) \, \ldots \, \vect{D}_{\boldsymbol{\phi}_j(t)} \vect{D}_{\boldsymbol{\phi}_j(\tau_B)}^{\Htran} x_{lk}^*(\tau_B) ] \} \\ \notag
& \!=  \vect{\Lambda}_{jlk} [  x_{lk}^*(\tau_1) e^{-\frac{\delta}{2} |t-\tau_1|} \vect{I}_N \, \ldots \,  x_{lk}^*(\tau_B) e^{-\frac{\delta}{2} |t-\tau_B|} \vect{I}_N ] \\ & \!= \tilde{\vect{x}}_{lk}^{\Htran}  \vect{D}_{\boldsymbol{\delta}(t)} \kron \vect{\Lambda}_{jlk} \label{eq:cross-correlation-LMMSE}
\end{align}
since $\mathbb{E}\{ e^{\imath \phi_{n,t_1}} e^{-\imath \phi_{n,t_2}} \} = e^{-\frac{\delta}{2} |t_1-t_2|}$ and by exploiting the fact that diagonal matrices commute.
Furthermore, we have that
\begin{align}
& \mathbb{E}\{ \boldsymbol{\psi}_j \boldsymbol{\psi}_j^{\Htran} \} \notag \\ &= \! \sum_{\ell=1}^{L} \sum_{m=1}^{K} \!
\mathbb{E}\Big\{ [   \vect{D}_{\boldsymbol{\phi}_j(\tau_1)}^{\Htran} x_{\ell m}^*(\tau_1) \, \ldots \, \vect{D}_{\boldsymbol{\phi}_j(\tau_B)}^{\Htran} x_{\ell m}^*(\tau_B) ]^{\Htran} \vect{h}_{j\ell m} \notag \\ & \times \vect{h}_{j\ell m}^{\Htran} [  \vect{D}_{\boldsymbol{\phi}_j(\tau_1)}^{\Htran} x_{\ell m}^*(\tau_1) \, \ldots \, \vect{D}_{\boldsymbol{\phi}_j(\tau_B)}^{\Htran} x_{\ell m}^*(\tau_B) ] \Big\} \notag \\ &+ \mathbb{E}\{ [ \boldsymbol{\upsilon}_j^{\Htran}(\tau_1)  \, \ldots \boldsymbol{\upsilon}_j^{\Htran}(\tau_B) ]^{\Htran} [ \boldsymbol{\upsilon}_j^{\Htran}(\tau_1)  \, \ldots \boldsymbol{\upsilon}_j^{\Htran}(\tau_B) ] \} \notag \\ &+ \mathbb{E}\{ [ \boldsymbol{\eta}_j^{\Htran}(\tau_1)  \, \ldots \boldsymbol{\eta}_j^{\Htran}(\tau_B) ]^{\Htran} [ \boldsymbol{\eta}_j^{\Htran}(\tau_1)  \, \ldots \boldsymbol{\eta}_j^{\Htran}(\tau_B) ] \} \notag \\
&= \underbrace{\sum_{\ell=1}^{L} \sum_{m=1}^{K} \Big( \vect{X}_{\ell m} \kron \vect{\Lambda}_{j \ell m} +
 \kappa^2 \vect{D}_{| \tilde{\vect{x}}_{\ell m}|^2} \kron \vect{\Lambda}_{j \ell m} \Big) + \xi \vect{I}_{BN}}_{\triangleq \boldsymbol{\Psi}_j}.
 \label{eq:correlation-LMMSE}
\end{align}
The LMMSE estimator in \eqref{eq:LMMSE-estimator} now follows from \eqref{eq:cross-correlation-LMMSE} and \eqref{eq:correlation-LMMSE}. The error covariance matrix in \eqref{eq:LMMSE-error-cov} is computed as $\vect{\Lambda}_{jlk} - \mathbb{E}\{ \vect{h}_{jlk}(t) \boldsymbol{\psi}_j^{\Htran} \} \left( \mathbb{E}\{ \boldsymbol{\psi}_j \boldsymbol{\psi}_j^{\Htran} \}  \right)^{-1} (\mathbb{E}\{ \vect{h}_{jlk}(t) \boldsymbol{\psi}_j^{\Htran} \})^{\Htran}$ \cite[Ch.~12]{Kay1993a}.

\section*{Appendix C: Proof of Lemma \ref{lemma:achievable-rates}}
\label{proof:lemma:achievable-rates}

Since the effective channels vary with $t$, we follow the approach in \cite{Pitarokoilis2014a} and compute one ergodic achievable rate for each $t \in \mathcal{D}$.
We obtain \eqref{eq:achievable-rate} by taking the average of these rates. The SINR in \eqref{eq:achievable-SINR} is obtained by treating the uncorrelated inter-user interference and distortion noise as independent Gaussian noise, which is a worst-case assumption when computing the mutual information \cite{Hassibi2003a}. In addition, we follow an approach from \cite{Medard2000a} and only exploit the knowledge of the average effective channel $\mathbb{E}\{ \vect{v}_{jk}^{\Htran}(t) \vect{h}_{jjk}(t) \}$  in the detection, while the deviation from the average effective channel is treated as worst-case Gaussian noise with variance $\mathbb{E}\{ |\vect{v}_{jk}^{\Htran}(t) \vect{h}_{jjk}(t)|^2 \} -|\mathbb{E}\{ \vect{v}_{jk}^{\Htran}(t) \vect{h}_{jjk}(t) \}|^2 $.

\section*{Appendix D: Proof of Theorem \ref{theorem:MRC-expectations}}
\label{proof:theorem:MRC-expectations}

The expressions in Theorem \ref{theorem:MRC-expectations} are derived one at the time. For brevity, we use the following notations in the derivations:
\begin{align}
\vect{A}_{jlk}(t) &= \left( \tilde{\vect{x}}_{lk}^{\Htran}  \vect{D}_{\boldsymbol{\delta}(t)} \kron \vect{\Lambda}_{jlk} \right) \boldsymbol{\Psi}^{-1}_j  \label{eq:A-matrix}\\
\vect{D}_{|\vect{h}_{jlk}|^2} &= \diag\bigg( |h_{jlk}^{(1)}|^2,\ldots, |h_{jlk}^{(N)}|^2\bigg) \\
\vect{B}_{jklm}(t) &= \vect{\Lambda}_{jlm} \vect{A}_{jjk}(t) \\
\vect{M}_{jklm}(t) & = \vect{D}_{\boldsymbol{\phi}_j(t)}^{\Htran} \vect{A}_{jjk}(t) \notag \\ & \,\,\, \times [ \vect{D}_{\boldsymbol{\phi}_j(\tau_1)}^{\Ttran} x_{lm}(\tau_1) \, \ldots \, \vect{D}_{\boldsymbol{\phi}_j(\tau_B)}^{\Ttran} x_{lm}(\tau_B) ]^{\Ttran}.
\end{align}

We begin with \eqref{eq:MRC-squared-norm} and exploit that $\vect{v}_{jk}(t) = \hat{\vect{h}}_{jlk}(t)$ is an LMMSE estimate to see that
\begin{equation} \label{eq:MRC-squared-norm-repeated}
\begin{split}
&\mathbb{E}\{ \| \vect{v}_{jk}(t) \|^2\} = \tr( \vect{\Lambda}_{jjk} - \vect{C}_{jjk}(t) ) \\ &= \tr \left( \left( \tilde{\vect{x}}_{jk}^{\Htran}  \vect{D}_{\boldsymbol{\delta}(t)} \kron \vect{\Lambda}_{jjk} \right) \boldsymbol{\Psi}^{-1}_j \left(  \vect{D}_{\boldsymbol{\delta}(t)}^{\Htran} \tilde{\vect{x}}_{jk}   \kron \vect{\Lambda}_{jjk} \right) \right)
\end{split}
\end{equation}
which proves \eqref{eq:MRC-squared-norm}. Next, we exploit that $\hat{\vect{h}}_{jjk}(t) = \vect{A}_{jjk}(t) \boldsymbol{\psi}_j$ and note that
\begin{equation} \label{eq:MRC-first-moment-derivation}
\begin{split}
&\mathbb{E}\{ \vect{v}_{jk}^{\Htran}(t) \vect{h}_{jjk}(t) \} = \tr \left(  \mathbb{E}\{ \vect{h}_{jjk}(t) \boldsymbol{\psi}_j^{\Htran} \}  \vect{A}_{jjk}^{\Htran}(t) \right) \\ & =
\tr \left(
\left( \tilde{\vect{x}}_{jk}^{\Htran}  \vect{D}_{\boldsymbol{\delta}(t)} \kron \vect{\Lambda}_{jjk} \right)  \vect{A}_{jjk}^{\Htran}(t) \right) \\
 & = \tr \left( \left( \tilde{\vect{x}}_{jk}^{\Htran}  \vect{D}_{\boldsymbol{\delta}(t)} \kron \vect{\Lambda}_{jjk} \right) \boldsymbol{\Psi}^{-1}_j \left(  \vect{D}_{\boldsymbol{\delta}(t)}^{\Htran} \tilde{\vect{x}}_{jk}   \kron \vect{\Lambda}_{jjk} \right) \right)
\end{split}
\end{equation}
where the second equality follows from \eqref{eq:cross-correlation-LMMSE} and the third equality follows from the full expression of $\vect{A}_{jjk}(t)$ in \eqref{eq:A-matrix}. Observe that the expression \eqref{eq:MRC-first-moment-derivation} is the same as for $\mathbb{E}\{ \| \vect{v}_{jk}(t) \|^2\}$ in \eqref{eq:MRC-squared-norm-repeated}.

Next, the second-order moment in \eqref{eq:MRC-second-moment} can be expanded as
\begin{equation} \label{eq:MRC-second-moment-derivation}
\begin{split}
&\mathbb{E}\left\{ | \vect{v}_{jk}^{\Htran}(t) \vect{h}_{jlm}(t) |^2 \right\} \\ &= \mathbb{E}\{ \tr( \vect{A}_{jjk}^{\Htran}(t) \vect{h}_{jlm}(t) \vect{h}_{jlm}^{\Htran}(t) \vect{A}_{jjk}(t) \boldsymbol{\psi}_j \boldsymbol{\psi}_j^{\Htran}  )  \} \\
&=  \tr \bigg( \vect{A}_{jjk}^{\Htran}(t) \vect{\Lambda}_{jlm}  \vect{A}_{jjk}(t) \Big(
\boldsymbol{\Psi}_j - \vect{X}_{l m} \kron \vect{\Lambda}_{jlm}
 \Big)  \bigg) \\
&+ \mathbb{E}\left\{ \tr( \vect{M}_{jklm}^{\Htran}(t) \vect{h}_{jlm} \vect{h}_{jlm}^{\Htran}  \vect{M}_{jklm}(t) \vect{h}_{jlm} \vect{h}_{jlm}^{\Htran} )  \right\}  \\
&+ \kappa^2 \mathbb{E}\left\{ \tr \big( \vect{A}_{jjk}^{\Htran}(t) \vect{h}_{jlm} \vect{h}_{jlm}^{\Htran}  \vect{A}_{jjk}(t) (\vect{D}_{| \tilde{\vect{x}}_{l m}|^2} \kron \vect{D}_{|\vect{h}_{jlm}|^2})  \big)  \right\} 
\end{split}
\end{equation}
where the first term follows from computing separate expectations for the parts of $\boldsymbol{\psi}_j \boldsymbol{\psi}_j^{\Htran}$ that are independent of $\vect{h}_{jlm}(t) \vect{h}_{jlm}^{\Htran}(t)$. The remaining two terms take care of the statistically dependent terms. The middle term is simplified as
\begin{equation} \label{eq:MRC-second-moment-derivation2}
\begin{split}
&\mathbb{E}\left\{ \tr( \vect{M}_{jklm}^{\Htran}(t) \vect{h}_{jlm} \vect{h}_{jlm}^{\Htran}  \vect{M}_{jklm}(t) \vect{h}_{jlm} \vect{h}_{jlm}^{\Htran} )  \right\}  \\ & =
 \mathbb{E}\left\{ |\tr( \vect{\Lambda}_{jlm} \vect{M}_{jklm}(t) )|^2 \right\}  \\ &+ \mathbb{E}\left\{ \tr( \vect{\Lambda}_{jlm} \vect{M}_{jklm}(t) \vect{\Lambda}_{jlm} \vect{M}_{jklm}^{\Htran}(t))\right\} 
\end{split}
\end{equation}
by computing the expectation with respect to $\vect{h}_{jlm}$ using Lemma \ref{lemma:second-order-moment-Gaussian} in Appendix A.
The first expectation in \eqref{eq:MRC-second-moment-derivation2} is now computed by expanding the expression as
\begin{equation} \label{eq:derivation:trM2}
\begin{split}
&\mathbb{E}\{ |\tr( \vect{\Lambda}_{jlm} \vect{M}_{jklm}(t) )|^2 \} = \mathbb{E}\Big\{ \Big|\tr \Big( \vect{B}_{jklm}(t) \\ &
 \times [ \vect{D}_{\boldsymbol{\phi}_j(\tau_1)}^{\Ttran} x_{lm}(\tau_1) \, \ldots \, \vect{D}_{\boldsymbol{\phi}_j(\tau_B)}^{\Ttran} x_{lm}(\tau_B) ]^{\Ttran} \vect{D}_{\boldsymbol{\phi}_j(t)}^{\Htran} \Big) \Big|^2 \Big\}
\\ &= \mathbb{E}\Big\{ \sum_{n_1=1}^{N} \sum_{b_1=1}^{B} [ \vect{B}_{jklm}(t) \vect{E}_{b_1}]_{n_1 n_1} x_{lm}(\tau_{b_1}) e^{\imath \phi_{n_1,\tau_{b_1}}} e^{- \imath \phi_{n_1,t}}
\\ & \quad \times \sum_{n_2=1}^{N}  \sum_{b_2=1}^{B}
[ \vect{E}_{b_2}^{\Htran}\vect{B}_{jklm}^{\Htran}(t) ]_{n_2 n_2} x_{lm}^* (\tau_{b_2}) e^{-\imath \phi_{n_2,\tau_{b_2}}} e^{ \imath \phi_{n_2,t}}
 \Big\} \\
& = \sum_{n_1,n_2,b_1,b_2} [ \vect{B}_{jklm}(t) \vect{E}_{b_1}]_{n_1 n_1} [ \vect{E}_{b_2}^{\Htran}\vect{B}_{jklm}^{\Htran}(t) ]_{n_2 n_2}  \\
 & \times x_{lm}(\tau_{b_1}) x_{lm}^* (\tau_{b_2}) \mathbb{E}\left\{ e^{\imath (\phi_{n_1,\tau_{b_1}}-  \phi_{n_1,t} -\phi_{n_2,\tau_{b_2}} + \phi_{n_2,t} ) } \right\}
\end{split}
\end{equation}
where $\vect{E}_{b_i} = \vect{e}_{b_i} \kron \vect{I}_N$ and $\vect{e}_{b_i} \in \mathbb{C}^{B \times 1}$ is the $b_i$th column of $\vect{I}_B$. The phase-drift expectation depends on the use of a CLO or SLOs:
\begin{equation}
\begin{split}
\mathbb{E} &\left\{ e^{\imath (\phi_{n_1,\tau_{b_1}}-  \phi_{n_1,t} -\phi_{n_2,\tau_{b_2}} + \phi_{n_2,t} ) } \right\} \\ &= \begin{cases}
e^{-\frac{\delta}{2} |\tau_{b_1}-\tau_{b_2}|}, & \textrm{if a CLO}, \\
e^{-\frac{\delta}{2} |\tau_{b_1}-\tau_{b_2}|}, & \textrm{if SLOs and } n_1 = n_2, \\
e^{-\frac{\delta}{2} |t-\tau_{b_1}|} e^{-\frac{\delta}{2} |t-\tau_{b_2}|}, & \textrm{if SLOs and } n_1 \neq n_2.
\end{cases}
\end{split}
\end{equation}
Since $x_{lm}(\tau_{b_1}) x_{lm}^* (\tau_{b_2}) e^{-\frac{\delta}{2} |\tau_{b_1}-\tau_{b_2}|} = [ \bar{\vect{X}}_{lm} ]_{b_1,b_2} = \vect{e}_{b_1}^{\Htran} \bar{\vect{X}}_{lm} \vect{e}_{b_2}$ in the case of a CLO, \eqref{eq:derivation:trM2} becomes
\begin{equation} \label{eq:derivation:trM3-CLO}
\begin{split}
& \sum_{n_1,n_2,b_1,b_2} [ \vect{B}_{jklm}(t) \vect{E}_{b_1}]_{n_1 n_1} [ \vect{E}_{b_2}^{\Htran}\vect{B}_{jklm}^{\Htran}(t) ]_{n_2 n_2}  \vect{e}_{b_1}^{\Htran} \bar{\vect{X}}_{lm} \vect{e}_{b_2} \\
& = \sum_{n_1,n_2} \vect{e}_{n_1}^{\Htran} \vect{B}_{jklm}(t) (\bar{\vect{X}}_{lm} \kron \vect{e}_{n_1} \vect{e}_{n_2}^{\Htran}) \vect{B}_{jklm}^{\Htran}(t) \vect{e}_{n_2}
\end{split}
\end{equation}
where $\vect{e}_{n} \in \mathbb{C}^{N \times 1}$ is the $n$th column of $\vect{I}_N$ (recall also the definitions of $\vect{E}_{b}$ and $\vect{e}_b$ above).

Next, we note that $e^{-\frac{\delta}{2} |t-\tau_{b}|}  x_{lm}(\tau_{b}) = [ \vect{D}_{\boldsymbol{\delta}(t)}^{\Htran} \tilde{\vect{x}}_{lm} ]_{b}$.
In the case of SLOs, \eqref{eq:derivation:trM2} then becomes
\begin{equation} \label{eq:derivation:trM3-SLO}
\begin{split}
&\condSum{n_1,n_2,b_1,b_2}{n_1 \neq n_2}{} [ \vect{B}_{jklm}(t) \vect{E}_{b_1}]_{n_1 n_1} [ \vect{E}_{b_2}^{\Htran}\vect{B}_{jklm}^{\Htran}(t) ]_{n_2 n_2}  \\
 & \quad \quad \times [ \vect{D}_{\boldsymbol{\delta}(t)}^{\Htran} \tilde{\vect{x}}_{lm} ]_{b_1} [  \tilde{\vect{x}}_{lm}^{\Htran} \vect{D}_{\boldsymbol{\delta}(t)} ]_{b_2} \\
 &+ \sum_{n,b_1,b_2} [ \vect{B}_{jklm}(t) \vect{E}_{b_1}]_{n n} [ \vect{E}_{b_2}^{\Htran}\vect{B}_{jklm}^{\Htran}(t) ]_{n n} [ \bar{\vect{X}}_{lm} ]_{b_1,b_2} \\
 &= \Big| \sum_{n,b} [ \vect{B}_{jklm}(t) \vect{E}_{b}]_{n n} [ \vect{D}_{\boldsymbol{\delta}(t)}^{\Htran} \tilde{\vect{x}}_{lm} ]_{b} \Big|^2 \\
 &+  \sum_{n,b_1,b_2} [ \vect{B}_{jklm}(t) \vect{E}_{b_1}]_{n n} [ \vect{E}_{b_2}^{\Htran}\vect{B}_{jklm}^{\Htran}(t) ]_{n n} \\ & \quad \quad \times[ \bar{\vect{X}}_{lm} - \vect{D}_{\boldsymbol{\delta}(t)}^{\Htran} \tilde{\vect{x}}_{lm}  \tilde{\vect{x}}_{lm}^{\Htran} \vect{D}_{\boldsymbol{\delta}(t)}]_{b_1,b_2} \\
 &= \big| \tr( \vect{A}_{jjk}(t) ( \vect{D}_{\boldsymbol{\delta}(t)}^{\Htran} \tilde{\vect{x}}_{lm} \kron \vect{\Lambda}_{jlm} )   ) \big|^2 + \sum_{n} \vect{e}_n^{\Htran} \vect{B}_{jklm}(t)\\
 & \times \big( (\bar{\vect{X}}_{lm} - \vect{D}_{\boldsymbol{\delta}(t)}^{\Htran} \tilde{\vect{x}}_{lm}  \tilde{\vect{x}}_{lm}^{\Htran} \vect{D}_{\boldsymbol{\delta}(t)}) \kron \vect{e}_n \vect{e}_n^{\Htran}  \big) \vect{B}_{jklm}^{\Htran}(t) \vect{e}_n.
\end{split}
\end{equation}
The second expectation in \eqref{eq:MRC-second-moment-derivation2} is computed along the same lines as in \eqref{eq:correlation-LMMSE} and becomes
\begin{equation}
\begin{split}
&\mathbb{E}\{ \tr( \vect{\Lambda}_{jlm} \vect{M}_{jklm}(t) \vect{\Lambda}_{jlm} \vect{M}_{jklm}^{\Htran}(t)) \} \\
&= \tr \big( \vect{\Lambda}_{jlm} \vect{A}_{jjk}(t) ( \bar{\vect{X}}_{lm} \kron \vect{\Lambda}_{jlm} ) \vect{A}_{jjk}^{\Htran}(t) \big).
\end{split}
\end{equation}

It remains to compute the last term in \eqref{eq:MRC-second-moment-derivation}.
We exploit the following expansion of diagonal matrices: $\vect{D}_{| \tilde{\vect{x}}_{l m}|^2} = \sum_{b=1}^{B}  |x_{l m}(\tau_b)|^2 \vect{e}_b \vect{e}_b^{\Htran}$ and
$\vect{D}_{|\vect{h}_{jlm}|^2} = \sum_{n=1}^{N} | \vect{e}_n^{\Htran} \vect{h}_{jlm}  |^2 \vect{e}_n \vect{e}_n^{\Htran}$, where $\vect{e}_b$ is the $b$th column of $\vect{I}_B$ and
$\vect{e}_n$ is the $n$th column of $\vect{I}_N$. 
Plugging this into the last term in \eqref{eq:MRC-second-moment-derivation} yields
\begin{equation} \label{eq:MRC-second-moment-derivation3}
\begin{split}
&\sum_{b,n} |x_{l m}(\tau_b)|^2 \mathbb{E}\left\{ |  \vect{h}_{jlm}^{\Htran}  \vect{A}_{jjk}(t) (\vect{e}_b \kron \vect{e}_n \vect{e}_n^{\Htran}) \vect{h}_{jlm} |^2 \right\} \\
& = \sum_{b,n} \left|x_{l m}(\tau_b)|^2  |\tr( \vect{\Lambda}_{jlm} \vect{A}_{jjk}(t) (\vect{e}_b \kron \vect{e}_n \vect{e}_n^{\Htran}) )\right|^2 \\
& + \sum_{b,n} |x_{l m}(\tau_b)|^2 \tr \big( \vect{\Lambda}_{jlm} \vect{A}_{jjk}(t) (\vect{e}_b \kron \vect{e}_n \vect{e}_n^{\Htran}) \\ & \quad \times \vect{\Lambda}_{jlm} (\vect{e}_b^{\Htran} \kron \vect{e}_n \vect{e}_n^{\Htran}) \vect{A}_{jjk}^{\Htran}(t) \big) \\
& = \sum_{n}   \vect{e}_n^{\Htran} \vect{\Lambda}_{jlm} \vect{A}_{jjk}(t) (\vect{D}_{| \tilde{\vect{x}}_{l m}|^2} \kron  \vect{e}_n \vect{e}_n^{\Htran})  \vect{A}_{jjk}^{\Htran}(t) \vect{\Lambda}_{jlm} \vect{e}_n \\
& +  \tr \big( \vect{\Lambda}_{jlm} \vect{A}_{jjk}(t) (\vect{D}_{| \tilde{\vect{x}}_{l m}|^2} \kron \vect{\Lambda}_{jlm} ) \vect{A}_{jjk}^{\Htran}(t) \big)
\end{split}
\end{equation}
where the first equality follows from Lemma \ref{lemma:second-order-moment-Gaussian} and the second equality from reverting the matrix expansions wherever possible. Plugging \eqref{eq:MRC-second-moment-derivation2}--\eqref{eq:MRC-second-moment-derivation3} into \eqref{eq:MRC-second-moment-derivation} and utilizing $\bar{\vect{X}}_{l m} + \kappa^2 \vect{D}_{|\tilde{\vect{x}}_{lm}|^2}  = \vect{X}_{l m}$, we obtain \eqref{eq:MRC-second-moment} by removing the special notation that was introduced in the beginning of this appendix.

Finally, we compute the expectation in \eqref{eq:MRC-cross-moment} by noting that
\begin{equation}\label{eq:MRC-cross-moment_derivation}
\begin{split}
&\mathbb{E}\{ |\vect{v}_{jk}^{\Htran}(t) \boldsymbol{\upsilon}_j(t) |^2  \} = \mathbb{E}\{ \tr( \vect{A}_{jjk}^{\Htran}(t) \vect{\Upsilon}_j(t) \vect{A}_{jjk}(t) \boldsymbol{\psi}_j \boldsymbol{\psi}_j^{\Htran} ) \}\\
&=  \kappa^2 \sum_{l=1}^{L} \sum_{m=1}^{K} p_{lm}   \\
&\times\bigg(\tr \Big( \vect{A}_{jjk}^{\Htran}(t) \vect{\Lambda}_{jlm}  \vect{A}_{jjk}(t) \big( \boldsymbol{\Psi}_j - \vect{X}_{l m} \kron \vect{\Lambda}_{jlm} \big)  \Big) \\
&+ \mathbb{E} \{ \tr ( \vect{M}_{jklm}^{\Htran}(t) \vect{D}_{|\vect{h}_{jlm}|^2} \vect{M}_{jklm}(t)  \vect{h}_{jlm} \vect{h}_{jlm}^{\Htran}  ) \} \\
&+ \kappa^2 \mathbb{E}\{ \tr \big( \vect{A}_{jjk}^{\Htran}(t) \vect{D}_{|\vect{h}_{jlm}|^2} \vect{A}_{jjk}(t) (\vect{D}_{| \tilde{\vect{x}}_{l m}|^2} \kron \vect{D}_{|\vect{h}_{jlm}|^2})  \big) \} \bigg).
\end{split}
\end{equation}
The first equality follows by taking the expectation with respect to $\boldsymbol{\upsilon}_j (t)$ for fixed channel realizations. The second equality follows by taking separate expectations with respect to the terms of $\vect{\Upsilon}_j = \kappa^2 \sum_{l=1}^{L} \sum_{m=1}^{K} p_{lm}  \vect{D}_{|\vect{h}_{jlm}|^2} $ and $\boldsymbol{\psi}_j \boldsymbol{\psi}_j^{\Htran} $ that are independent. These give the first term in \eqref{eq:MRC-cross-moment_derivation} while the last two terms take care of the statistically dependent terms.

The expectation in the middle term of \eqref{eq:MRC-cross-moment_derivation} is computed as
\begin{equation} \label{eq:MRC-cross-moment_derivation2}
\begin{split}
&\mathbb{E} \{ \tr ( \vect{M}_{jklm}^{\Htran}(t) \vect{D}_{|\vect{h}_{jlm}|^2} \vect{M}_{jklm}(t)  \vect{h}_{jlm} \vect{h}_{jlm}^{\Htran}  ) \} \\
& = \sum_{n} \mathbb{E} \left\{ | \vect{h}_{jlm}^{\Htran} \vect{M}_{jklm}^{\Htran}(t) \vect{e}_n \vect{e}_n^{\Htran} \vect{h}_{jlm} |^2 \right\} \\
& = \sum_{n} \mathbb{E} \left\{ |   \vect{e}_n^{\Htran} \vect{\Lambda}_{jlm} \vect{M}_{jklm}^{\Htran}(t) \vect{e}_n  |^2\right\} \\
& + \sum_{n} \mathbb{E} \left\{   \vect{e}_n^{\Htran} \vect{\Lambda}_{jlm} \vect{e}_n  \vect{e}_n^{\Htran} \vect{M}_{jklm}(t) \vect{\Lambda}_{jlm}  \vect{M}_{jklm}^{\Htran}(t) \vect{e}_n  \right\} \\
& = \tr \big( \vect{\Lambda}_{jlm} \vect{A}_{jjk}(t) ( \bar{\vect{X}}_{l m}  \kron \vect{\Lambda}_{jlm} ) \vect{A}_{jjk}^{\Htran}(t) \big)
\\
& + \sum_{n}  \vect{e}_n^{\Htran} \vect{\Lambda}_{jlm} \vect{A}_{jjk}(t) ( \bar{\vect{X}}_{l m}  \kron \vect{e}_n \vect{e}_n^{\Htran} ) \vect{A}_{jjk}^{\Htran}(t) \vect{\Lambda}_{jlm} \vect{e}_n
\end{split}
\end{equation}
where the first equality follows from the same diagonal matrix expansion as in \eqref{eq:MRC-second-moment-derivation3}, the second equality is due to Lemma \ref{lemma:second-order-moment-Gaussian} (and that diagonal matrices commute), and the third equality follows from computing the expectation with respect to phase-drifts as in  \eqref{eq:correlation-LMMSE} and then reverting the matrix expansions wherever possible.

Similarly, we have
\begin{equation} \label{eq:MRC-cross-moment_derivation3}
\begin{split}
&\mathbb{E}\{ \tr \big( \vect{A}_{jjk}^{\Htran}(t) \vect{D}_{|\vect{h}_{jlm}|^2} \vect{A}_{jjk}(t) (\vect{D}_{| \tilde{\vect{x}}_{l m}|^2} \kron \vect{D}_{|\vect{h}_{jlm}|^2})  \big) \} \\ =
\!&\! \sum_{n_1,n_2,b} |x_{l m}(\tau_b)|^2 \mathbb{E}\left\{ |  \vect{h}_{jlm}^{\Htran} \vect{e}_{n_1} \vect{e}_{n_1}^{\Htran} \vect{A}_{jjk}(t) ( \vect{e}_{b} \kron \vect{e}_{n_2} \vect{e}_{n_2}^{\Htran}) \vect{h}_{jlm} |^2 \right\} \\
=&  \sum_{n_1,n_2,b} |x_{l m}(\tau_b)|^2 \Big( \big| \tr \big( \vect{\Lambda}_{jlm} \vect{e}_{n_1} \vect{e}_{n_1}^{\Htran} \vect{A}_{jjk}(t) ( \vect{e}_{b} \kron \vect{e}_{n_2} \vect{e}_{n_2}^{\Htran})  \big) \big|^2 \\ & + \tr \big( \vect{\Lambda}_{jlm} \vect{e}_{n_1} \vect{e}_{n_1}^{\Htran} \vect{A}_{jjk}(t) ( \vect{e}_{b} \vect{e}_{b}^{\Htran}  \kron \vect{e}_{n_2} \vect{e}_{n_2}^{\Htran} \vect{\Lambda}_{jlm} ) \vect{A}_{jjk}^{\Htran}(t)  \big)  \Big) \\
=& \sum_{n_1} \vect{e}_{n_1}^{\Htran} \vect{\Lambda}_{jlm} \vect{A}_{jjk}(t) ( \vect{D}_{| \tilde{\vect{x}}_{l m}|^2} \kron \vect{e}_{n_1} \vect{e}_{n_1}^{\Htran} ) \vect{A}_{jjk}^{\Htran}(t) \vect{\Lambda}_{jlm} \vect{e}_{n_1} \\
& +  \tr \big( \vect{\Lambda}_{jlm} \vect{A}_{jjk}(t) ( \vect{D}_{| \tilde{\vect{x}}_{l m}|^2}  \kron \vect{\Lambda}_{jlm} ) \vect{A}_{jjk}^{\Htran}(t)  \big)
\end{split}
\end{equation}
where the first equality follows from the same diagonal matrix expansions as above, the second equality follows from Lemma \ref{lemma:second-order-moment-Gaussian} (and that diagonal matrices commute), and the third equality from reverting the matrix expansions wherever possible.

By plugging \eqref{eq:MRC-cross-moment_derivation2}  and \eqref{eq:MRC-cross-moment_derivation3}  into \eqref{eq:MRC-cross-moment_derivation} and utilizing $\bar{\vect{X}}_{l m} + \kappa^2 \vect{D}_{|\tilde{\vect{x}}_{lm}|^2}  = \vect{X}_{l m}$, we finally obtain \eqref{eq:MRC-cross-moment}.

\section*{Appendix E: Proof of Corollary \ref{corollary:asymptotic-SINR}}
\label{proof:corollary:asymptotic-SINR}

This corollary is obtained by dividing all the terms in $\mathrm{SINR}_{jk}(t)$ by $\frac{N^2}{A^2}$ and inspecting the scaling behavior as $N \rightarrow \infty$. Using the expressions in Theorem \ref{theorem:MRC-expectations} and utilizing that $\boldsymbol{\Psi}^{-1}_j = \widetilde{\boldsymbol{\Psi}}^{-1}_j \kron \vect{I}_{\frac{N}{A}}$, we observe that
$\frac{\xi A^2}{  N^2} \mathbb{E}\{ \| \vect{v}_{jk}(t) \|^2\}  = \frac{\xi A^2}{  N^2}  \tr ( \vect{F} \kron \vect{I}_{\frac{N}{A}} )= \frac{\xi A}{  N} \tr ( \vect{F} ) = \mathcal{O}(\frac{1}{N})$, where $\vect{F} = \left( \tilde{\vect{x}}_{jk}^{\Htran}  \vect{D}_{\boldsymbol{\delta}(t)} \!\kron\! \tilde{\vect{\Lambda}}_{jjk}^{(A)} \right) \widetilde{\boldsymbol{\Psi}}^{-1}_j \left(  \vect{D}_{\boldsymbol{\delta}(t)} \tilde{\vect{x}}_{jk}  \!\kron\! \tilde{\vect{\Lambda}}_{jjk}^{(A)} \right)$. Similarly, it is straightforward but lengthy to prove that
$\frac{A^2}{N^2} \mathbb{E}\{ |\vect{v}_{jk}^{\Htran}(t) \boldsymbol{\upsilon}_j(t) |^2  \} = \mathcal{O}(\frac{1}{N})$. The only terms in the SINR that remain as $N \rightarrow \infty$ are $\frac{A^2}{  N^2} (\mathbb{E}\{ \| \vect{v}_{jk}(t) \|^2\})^2 = \mathrm{Sig}_{jk}$ and
$\frac{A^2}{  N^2}  \mathbb{E}\{ | \vect{v}_{jk}^{\Htran}(t) \vect{h}_{jlm}(t) |^2 \} = \mathrm{Int}_{jklm} + \mathcal{O}(\frac{1}{N})$.

\section*{Appendix F: Proof of Corollary \ref{cor:scaling-law}}
\label{proof:cor:scaling-law}

The first step of the proof is to substitute the new parameters into the SINR expression in \eqref{eq:achievable-SINR} and scale all terms by $1/N^{1+z_3 \delta_{0} \min_{\tau} |t-\tau | }$. Since the distortion noise and receiver noise terms normally behave as $\mathcal{O}(N)$, it is straightforward (but lengthy) to verify that the (scaled) distortion noise and receiver noise terms go to zero when $N \rightarrow \infty$. Similarly, the signal term in the numerator which normally behave as $\mathcal{O}(N^2)$, will after the scaling behave as $\mathcal{O}(N^{1 - 2 \max(z_1,z_2) - z_3 \delta_{0} \min_{\tau} |t-\tau |})$. In the case of SLOs, the second-order interference moments $\mathbb{E}\{ | \vect{v}_{jk}^{\Htran}(t) \vect{h}_{jlm}(t) |^2 \}$ in the denominator exhibit the same scaling as the signal term. The scaling law in \eqref{eq:scaling-law} then follows from that we want the signal and interference terms to be non-vanishing in the asymptotic limit; that is, $1 - 2 \max(z_1,z_2) - z_3 \delta_{0} \min_{\tau} |t-\tau | > 1$. In the case of a CLO, the second-order interference moments behave as $\mathcal{O}(N^{1 - 2 \max(z_1,z_2) })$ and do not depend on $z_3$. To make the signal and interference terms have the same scaling and be non-vanishing, we thus need to set $z_3=0$ and $\max(z_1,z_2) \leq \frac{1}{2}$.

\bibliographystyle{IEEEbib}
\bibliography{IEEEabrv,refs}

\end{document}